\documentclass[journal]{IEEEtran}
\usepackage{graphicx}
\usepackage{cite}
\usepackage{array}
\usepackage{amsmath,amssymb,amsfonts}
\usepackage{amsthm}
\usepackage{algorithm}
\usepackage{algorithmicx}
\usepackage{algpseudocode}
\usepackage{enumitem}
\usepackage{pifont}
\usepackage{subfig}
\allowdisplaybreaks[3]
\makeatletter
\newtheoremstyle{lem}% name
{5pt}%      Space above, empty = `usual value'
{5pt}%      Space below
{}% Body font
{}%         Indent amount (empty = no indent, \parindent = para indent)
{\bfseries}% Thm head font
{:}%        Punctuation after thm head
{.2em}%     Space after thm head: " " = normal interword space;
{}% Thm head spec

\theoremstyle{lem}
\newtheorem{myLem}{Lemma}
\newtheoremstyle{pro}% name
{0pt}%      Space above, empty = `usual value'
{0pt}%      Space below
{}% Body font
{}%         Indent amount (empty = no indent, \parindent = para indent)
{\bfseries\itshape}% Thm head font
{:}%        Punctuation after thm head
{.2em}%     Space after thm head: " " = normal interword space;
{}% Thm head spec

\theoremstyle{pro}
\newtheorem*{myPro}{Proof}
\newtheoremstyle{def}% name
{5pt}%      Space above, empty = `usual value'
{5pt}%      Space below
{\itshape}% Body font
{}%         Indent amount (empty = no indent, \parindent = para indent)
{\bfseries}% Thm head font
{.}%        Punctuation after thm head
{.2em}%     Space after thm head: " " = normal interword space;
{}% Thm head spec

\theoremstyle{def}
\newtheorem{myDef}{Definition}% all text supplied in the note

\newtheoremstyle{rem}% name
{5pt}%      Space above, empty = `usual value'
{5pt}%      Space below
{}% Body font
{}%         Indent amount (empty = no indent, \parindent = para indent)
{\bfseries}% Thm head font
{.}%        Punctuation after thm head
{.2em}%     Space after thm head: " " = normal interword space;
{}% Thm head spec
\theoremstyle{rem}
\newtheorem{myRem}{Remark}

\newcommand{\Rmnum}[1]{\expandafter\@slowromancap\romannumeral #1@}
\renewcommand\normalsize{%
	\@setfontsize\normalsize\@xpt\@xiipt
	\abovedisplayskip 5\p@ \@plus2\p@ \@minus5\p@
	\abovedisplayshortskip \z@ \@plus3\p@
	\belowdisplayshortskip 6\p@ \@plus3\p@ \@minus5\p@
	\belowdisplayskip \abovedisplayskip
	\let\@listi\@listI}
\renewcommand*\env@matrix[1][\arraystretch]{%
	\edef\arraystretch{#1}%
	\hskip -\arraycolsep
	\let\@ifnextchar\new@ifnextchar
	\array{*\c@MaxMatrixCols c}}
\makeatother
\ifCLASSINFOpdf
\else
\fi
\begin{document}

%
% paper title
% Titles are generally capitalized except for words such as a, an, and, as,
% at, but, by, for, in, nor, of, on, or, the, to and up, which are usually
% not capitalized unless they are the first or last word of the title.
% Linebreaks \\ can be used within to get better formatting as desired.
% Do not put math or special symbols in the title.
\title{JOET: Sustainable Vehicle-assisted Edge \\ Computing for Internet of Vehicles}
%
%
% author names and IEEE memberships
% note positions of commas and nonbreaking spaces ( ~ ) LaTeX will not break
% a structure at a ~ so this keeps an author's name from being broken across
% two lines.
% use \thanks{} to gain access to the first footnote area
% a separate \thanks must be used for each paragraph as LaTeX2e's \thanks
% was not built to handle multiple paragraphs
%

\author{Wei~Huang,
        Neal~N.~Xiong,~\IEEEmembership{Senior~Member,~IEEE,}
        Shahid~Mumtaz,~\IEEEmembership{Senior~Member,~IEEE}% <-this % stops a space
\thanks{This work was supported by the National Natural Science Foundation of China (62072475, 61772554) (*Corresponding author: Neal N. Xiong).}
\thanks{W. Huang is with School of Computer Science and Engineering, Central South University, ChangSha 410083 China. (e-mail:csu\_hw@csu.edu.cn).}
\thanks{N. N. Xiong is with the Department of Mathematics and Computer Science, Northeastern State University, OK, 74464, USA. (e-mail: xiongnaixue@gmail.com).}
\thanks{S. Mumtaz is with the Instituto de Telecomunicac$\c{c}\tilde{o}$es, Portugal, 1049-001 Aveiro, Portugal. (e-mail: smumtaz@av.it.pt).}}
\maketitle

% As a general rule, do not put math, special symbols or citations
% in the abstract or keywords.
\begin{abstract}
Task offloading in Internet of Vehicles (IoV) involves numerous steps and optimization variables such as: where to offload tasks, how to allocate computation resources, how to adjust offloading ratio and transmit power for offloading, and such optimization variables and hybrid combination features are highly coupled with each other. Thus, this is a fully challenge issue to optimize these variables for task offloading to sustainably reduce energy consumption with load balancing while ensuring that a task is completed before its deadline. In this paper, we first provide a Mixed Integer Nonlinear Programming Problem (MINLP) formulation for such task offloading under energy and deadline constraints in IoV. Furthermore, in order to efficiently solve the formulated MINLP, we decompose it into two subproblems, and design a low-complexity Joint Optimization for Energy Consumption and Task Processing Delay (JOET) algorithm to optimize selection decisions, resource allocation, offloading ratio and transmit power adjustment. We carry out extensive simulation experiments to validate JOET. Simulation results demonstrate that JOET outperforms many representative existing approaches in quickly converge and effectively reduce energy consumption and delay. Specifically, average energy consumption and task processing delay have been reduced by 15.93\% and 15.78\%, respectively, and load balancing efficiency has increased by 10.20\%.
\end{abstract}

% Note that keywords are not normally used for peerreview papers.

% For peer review papers, you can put extra information on the cover
% page as needed:
% \ifCLASSOPTIONpeerreview
% \begin{center} \bfseries EDICS Category: 3-BBND \end{center}
% \fi
%
% For peerreview papers, this IEEEtran command inserts a page break and
% creates the second title. It will be ignored for other modes.
\IEEEpeerreviewmaketitle

\section{INTRODUCTION}
% The very first letter is a 2 line initial drop letter followed
% by the rest of the first word in caps.
% 
% form to use if the first word consists of a single letter:
% \IEEEPARstart{A}{demo} file is ....
% 
% form to use if you need the single drop letter followed by
% normal text (unknown if ever used by the IEEE):
% \IEEEPARstart{A}{}demo file is ....
% 
% Some journals put the first two words in caps:
% \IEEEPARstart{T}{his demo} file is ....
% 
% Here we have the typical use of a "T" for an initial drop letter
% and "HIS" in caps to complete the first word.
\IEEEPARstart{T}{he} Internet of Vehicles (IoV) has gained more attentions for researchers than ever before \cite{pokhrel2020improving, vasudev2020lightweight, zhu2020multiagent}. On the one hand, the number of vehicles has grown rapidly which can provide more service \cite{marr2018much,teng2021low}. On the other hand, more and more Internet of Things Devices (IoTDs) are deployed in demand to sense data \cite{cisco2020cisco,zhu2020deep}, communicate and perform tasks collaboratively, thus promoting the development of IoV based applications \cite{cisco2020cisco,lueth2018state}. According to the IoT Analytics, the number of IoTDs worldwide is expected to reach 22 billion by 2025 \cite{lueth2018state}. In order to make these massive, hardware-simple, and short communication range IoTDs can be cost-effective to connect to the Internet. Employing Mobile Vehicles (MVs) as the relay to connect IoTDs to the Internet is a low-cost and effective method in which when MVs enter communication range of IoTDs \cite{huang2020result}, IoTDs can upload their data or tasks to the Internet via MVs \cite{saleem2020latency,huang2020intelligent}. Especially in smart city, such way has a huge sustainability advantage because IoTDs can access the Internet conveniently and at low cost due to the large scale and wide distribution of vehicles \cite{bonola2016opportunistic}.

Reviewing previous researches, task offloading in IoV mainly research offloading tasks of vehicles to Edge Servers (ESs). However, the widely deployed IoTDs in smart cities have a more urgent need for task offloading \cite{shen2021attdc,zhan2020mobility,zhou2017robust}. Relatively speaking, MVs have stronger computing power and communication capabilities, so they can upload their tasks to ESs at any time during the movement for offloading \cite{huang2020result,liu2020adaptive}. Once deployed, IoTDs cannot be moved like MVs. Unless there are ESs available for offloading within communication range of IoTDs, tasks from IoTDs can only be offloaded through MVs as the relay node \cite{huang2020result,liu2020adaptive}. These IoTDs can be deployed in areas that need to be monitored, such as Supervisory Control And Data Acquisition (SCADA) systems, so they have broad application prospects \cite{huang2020bd}. Therefore, offloading tasks from a huge number of IoTDs is more challenging and more valuable than the aforementioned task offloading of vehicles.

However, task offloading in such semi-connected networks which is studied in this paper is more challenging due to follow reasons.

(1) First, task offloading for IoTDs involves more steps and decisions, so the problem is more complicated. Tasks of IoTDs needs to be uploaded to MVs, and the decision that the MV needs to make after receiving the task is to complete the task by itself, or offload it to the ES. Not only that, in each step, we need to make a decision for the participants in task offloading. These decisions mainly include whether the task is to be offloaded? What is the ratio of offloaded the task? How much transmit power is applied? What are resources allocated to the task? Since each step affects the subsequent decisions, and each decision affects each other, the problem to be solved in this article is very complicated.

(2) Solving the optimization problem is quite challenging. The task offloading goals of IoTDs are as follow: (a) Time to complete the task is as small as possible; (b) Energy consumption of the network is as small as possible; (c) Network load balancing. Time to complete the task includes communication delay and computation delay. Then, resources allocated to the task are limited by the following factors. The first is resources of the task executor itself. If the task executor has more resources, the resources that can be allocated are more, and computation delay is smaller. Secondly, the network needs to balance the load of servers. Otherwise, it will lead to the limited resources that can be allocated to the task, which increases the task processing delay \cite{liu2019dynamic}.

It can be concluded that the goals of IoTDs task offloading in this article are to reduce task processing delay and total energy consumption with load balancing while ensuring task completion. Limitations are as follows: First, it is used to ensure that tasks can only be offloaded to one place; and the allocated resources need to meet the minimum task processing delay; and the sum of the total resources allocated to multiple tasks by an edge server cannot exceed the total amount of resources of the server; Finally, ratio of the offloaded computation bits does not exceed 1. Thus, offloading a task requires optimizing selection decisions, resource allocation, offloading ratio and transmit power adjustment. There is a highly complex coupling between optimization variables and hybrid combination features, so it is extremely challenging to solve such the task offloading problem. This paper formulates the above task offloading problem as a Mixed Integer Non-Linear Programming (MINLP) problem to better solve this problem. In summary, the main contributions of this paper are summarized as follows:

\vspace{1pt}
\hangindent 2.05em
$\bullet$ We first propose the task offloading issue for IoTDs through MVs in IoV. Unlike previous studies that only studied task offloading on MVs, we have raised the issue of how to sustainably offload tasks from IoTDs through MVs. We further formulate the problem of minimizing energy consumption and task processing delay as a system utility maximization problem, and then transform the problem into a MINLP with consideration of load balancing. 

\vspace{1pt}
\hangindent 2.05em	
$\bullet$ We propose a Joint Optimization for Energy consumption and Task processing delay (JOET) algorithm to effectively solve the proposed MINLP to minimize energy consumption and task processing delay. Our proposed JOET first determines selection decisions under given resource allocation, offloading ratio and transmit power, and then optimizes the three variables under obtained selection decisions. The optimization solution with less overhead can be found by repeating the above optimization steps until convergence. 

\vspace{1pt}
\hangindent 2.05em	
$\bullet$ We carry out extensive simulation experiments to validate our proposed JOET. Simulation results demonstrate that JOET outperforms many representative existing algorithms in quickly converge and effectively reduce energy consumption and delay. Specifically, the average energy consumption and task processing delay have been reduced by 15.93\% and 15.78\%, respectively, and the load balancing efficiency has increased by 10.20\%.

\vspace{1pt}
The rest of the paper is organized as follows. In Section \Rmnum{2}, we review related works. Section \Rmnum{3} introduces the system  model and problem description, while Section \Rmnum{4} presents the problem decoupling and the solution. We present the performance analysis in Section \Rmnum{5}. Finally, Section \Rmnum{6} concludes this paper and introduce future work. 

\section{RELATED WORKS}
Task offloading is to solve the lack of computing power of some devices in the current network, and computation resources of other devices are sufficient and available to complement each other, so as to make full use of network resources \cite{huang2020result,shiri2020communication}. On the one hand, for IoTDs with insufficient computing power such as sensor nodes and phones, MVs cannot meet the needs of latency-intensive tasks \cite{zhang2020hybrid,zhou2018direction}. On the other hand, a large number of Edge Servers (ESs) with powerful computing capabilities are deployed at the edge of the network, which can process tasks from the edge network nearby. There are a lot of studies focused on task offloading \cite{huang2020result,liu2020adaptive,liu2019dynamic,yu2016joint}. The task offloading mainly involves the following aspects: (1) where to offload \cite{huang2020result}; (2) Resource allocation \cite{liu2019dynamic,yu2016joint}; (3) Transmission rate adjustment \cite{liu2020adaptive}. 

The issue for where to offload refers to where tasks is generated, and where they are offloaded for computing. There are many devices that generate tasks, and offload strategies of tasks generated by different devices are different. The most studied among them is that IoTDs in the network edge can directly communicate with the Internet for task offloading. In such the task offloading, the network is divided into 3 layers \cite{yu2020intelligent}: (a) Things Layer, (b) Fog Layer, (c) Cloud. The task is generated by IoTDs in the things layer. Due to the relatively weak  computing power of IoTDs, they cannot meet the task requirements, and the task needs to be offload to Fog layer or cloud with more computation resources \cite{yu2020intelligent}. In such three-layer task offloading, some researchers have proposed game-based task offloading strategies, whose goal is to make these three layers maintain a more balanced load through game to maximize the utilization of network resources \cite{shiri2020communication,yu2020intelligent}. Further, Yu \textit{et al.} \cite{yu2020intelligent} think that different tasks have different computation requirements. Therefore, a task offloading strategy proposed by Yu \textit{et al.} \cite{yu2020intelligent} is: For tasks with a large amount of data, the near-computing strategy should be used to  save energy consumption caused by uploading tasks. For tasks with low computing power and less data, they can be uploaded to cloud computing to meet their computing power, and the energy paid by upload tasks is not much, so it can achieve better performance \cite{yu2020intelligent}. A task offloading scheme with load balancing proposed by Khafajiy \textit{et al.} \cite{al2020comitment}, where ESs of the Fog layer form a service group to provide services to the outside world. If an ES has a heavier load, it can also redistribute its own tasks to the lightly loaded ESs to achieve load balancing.

Resource allocation is also one of important research content. Generally, there may be multiple tasks that need to be executed on ESs \cite{yu2016joint,xu2019minimizing}. Executing tasks requires allocating computation resources such as CPU, memory, etc. Therefore, if more resources are allocated to a task, the time required to complete the task is smaller. However, since resources are limited and competitive, a good resource allocation can effectively improve resource utilization on the one hand, and shorten task processing time on the other hand. From the perspective of task scheduling, an ES is a resource body with multiple resources, so which ES to offload tasks belongs to a kind of resource scheduling with large granularity. From the perspective of fine-grained resource scheduling, Xu \textit{et al.} \cite{xu2019minimizing} proposed a resource scheduling strategy based on matching theory. They believe that some tasks requiring more computing power should be allocated more CPU resources while others tasks require more storage and communication resources. Thus, using matching theory to analyze computation resources in their method can achieve good results \cite{xu2019minimizing}.

Thirdly, different transmission rates have an important impact on task completion. If a task is not offloaded, completion time is computation delay executed locally. If the task is offloaded, communication delay of task offloading to ESs will be increased. Therefore, if a communication way with high speed is selected, communication delay can be reduced \cite{liu2020adaptive}. However, the communication way is mostly determined by the hardware. For example, in the research of Liu \textit{et al.} \cite{liu2020adaptive}, the transmission speed using Vehicle to Infrastructure (V2I) is higher, so transmission rate is small. On the other hand, transmission rate is related to transmit power used during communication. Generally, if a larger transmit power is used, transmission rate is larger  so that communication delay is smaller, but the communication-related energy consumption is usually increased \cite{guo2019uav}. Therefore, selection of transmit power becomes an important part of task offloading \cite{guo2019uav}. 

%Specifically, for IoTDs that cannot communicate directly with the Internet, there are also some researches on task offloading of resource-limited IoTDs. Guo \textit{et al.} \cite{guo2019uav} employed the method of dispatching Unmanned Aerial Vehicles (UAVs) to offload tasks. When UAVs flying in the sky above IoTDs, those tasks with computing needs will be offloaded to the UAVs carrying a powerful computing server \cite{guo2019uav}. Due to the limited energy of IoTDs and UAVs, Guo \textit{et al.} \cite{guo2019uav} studied how to allocate transmitted bits in both uplink and downlink and design a UAV trajectory to reduce the energy. 

It is worth noting that although the above optimization methods and strategies have been more researched. However, the task offloading issue in this paper for IoTDs through MVs in IoV is more complicated than previous studies. The task offloading studied in this paper includes task transmission from IoTDs to MVs, and then MVs decide whether to offloading to ESs. Finally, after tasks are completed, the result need to be returned from MVs or ESs to IoTDs. Not only the task offloading path is long, but there are also optimization variables such as selection decisions, resource allocation, offloading ratio and transmit power. These optimization variables make task offloading need to meet task processing delay and resource constraints, so it is comparatively challenging.

\section{SYSTEM MODEL AND DEFINITIONS}
\subsection{The vehicle-assisted VEC network}
Considering a vehicle-assisted VEC network composed of a set of Road Side Units $\mathcal{M}=\left\{1,\;\cdots,\;M\right\}$ and $K$ Internet of Things (IoTs) denoted by $\mathcal{K}=\left\{1,\;\cdots,\;K\right\}$ as shown in Fig. 1, where tasks can not only be computed locally, but also can be offloaded to vehicles for computing. Due to the limited computation resources, vehicles also act a relay to help computation bits of IoTDs transmit to RSUs equipped with a VEC server for offloading. Specifically, let $\varrho_{k}\left(0\leq \varrho_{k} \leq 1 \right)$ be offloading ratio of the task from IoTD $k$, that is a ratio of the offloaded computation bits to the total computation bits. In the network, under different $\varrho_k$, there are three computation strategies: Strategy \Rmnum{1}: All computation bits of the task from IoTD $k$ are computed locally; Strategy \Rmnum{2}: $(1-\varrho_k)$ computation bits are computed locally while $\varrho_k$ computation bits are offloaded to the vehicle; Strategy \Rmnum{3}: $(1-\varrho_k)$ computation bits are computed locally while $\varrho_k$ computation bits are offloaded to the server. Specifically (as shown in Fig. 1), when $\varrho_k=0$, IoTD $k$ only adopts Strategy \Rmnum{1}; when $ 0<\varrho_k <1$, IoTD $k$ adopts the both Strategy \Rmnum{1} and Strategy \Rmnum{2} or the both Strategy \Rmnum{1} and Strategy \Rmnum{3}; when $\varrho_k=1$, IoTD $k$ adopts Strategy \Rmnum{2} or Strategy \Rmnum{3}: All computation bits of the task are offloaded to the vehicle or the server. Each IoTD generates a computation task in each time slot \cite{zhang2019joint}, such as numerical calculation, data analysis, image processing, etc. For convenience, we use sufficiently constant $\delta_t$ to divide the period $T$ into $N$ slots with equal size, which are given by a set $\mathcal{N} = \left\{ 1,\;\cdots,\;N \right\}$. Specifically, the characteristics of a computation task can be represented by a 5-tuple parameter $u_{k}=<I_{k},~O_{k},~\tilde{c}_{k},~E_{k}^{max},~T_{k}^{max}>$, $k \in \mathcal{K}=\left\{1,\;\cdots,\;K\right\}$ where $I_k$ and $O_k$ indicate input and output data size of the task in terms of bits. $\tilde{c}_{k}$ represents the required computation resources for completing the task. $E^{max}_{k}$ and $T^{max}_{k}$ stands for the maximum energy consumption and the maximum permissible latency for offloading the task  $u_k$ and returning a result, respectively. Vehicles running on the road collect tasks from nearby IoTDs. When a task is completed, its results are also returned to the corresponding IoTD via a vehicle. Note that, the vehicle does not need to follow a specific trajectory during the whole process. Ref. \cite{bonola2016opportunistic} summarized the probability that a taxi's random waypoint can cover the entire city area within half an hour and at least one taxi in each area within 10 minutes. And the taxis are only a small part of urban vehicles. Dynamics and randomness of vehicles make the vehicle-assisted VEC extremely complex. In order to simplify the progress, we assume that when a IoTD has a demand to upload a task or receive a result, there will always be vehicles passing the IoTD based on Ref. \cite{bonola2016opportunistic}. We will focus on dynamic arrival of vehicles in our future research work. Every IoTD, vehicle and VEC server is assumed equipped with one single antenna \cite{zhang2019joint}. 
\begin{figure}[!tbp]
	\centering
	\includegraphics[width=0.96\linewidth]{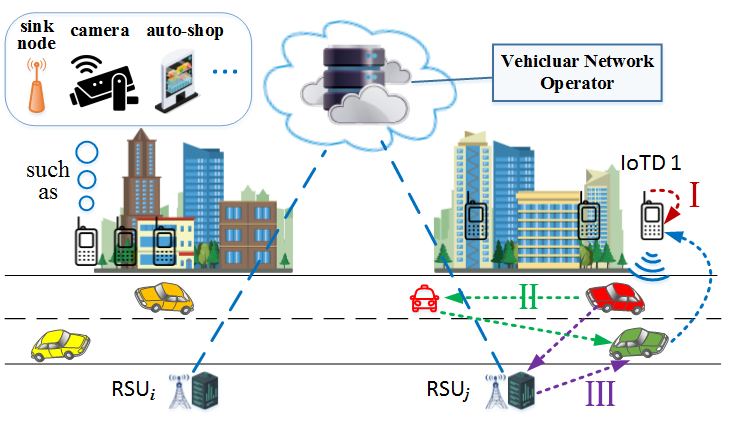}
	\caption{\label{fig1} The vehicle-assisted VEC network.}
\end{figure}

Considering that the life of IoTDs is constrained by energy and sensitive latency, each $u_k$ can reach the vehicle via wireless channel for either computing or relaying. In particular, where tasks are offloaded is determined based on maximizing system utility. In other words, at the beginning of each time slot, Vehicular Network Operator (VNO) collects basic information about tasks from IoTDs. Meanwhile, the VNO also updates a resource allocation table (RAT) \cite{ndikumana2019joint}, the RAT records the available resources of each VEC server. Then the VNO determines a computation strategy for each IoTD by the proposed JOET algorithm. Specifically, if $u_k$ is offloaded to the server, then computation resources allocated to $u_k$ are more than resources provided by the vehicle to make up for the long-distance cost. The VEC servers all have the same initial resources $F[0]$. For RSU $m$, the free computation resources $F_m[n]$ at time slot $n$ is equal to the sum of the remaining resources and the released resources at time slot $n-1$. 
\begin{figure}[!tbp]
	\centering
	\includegraphics[width=0.96\linewidth]{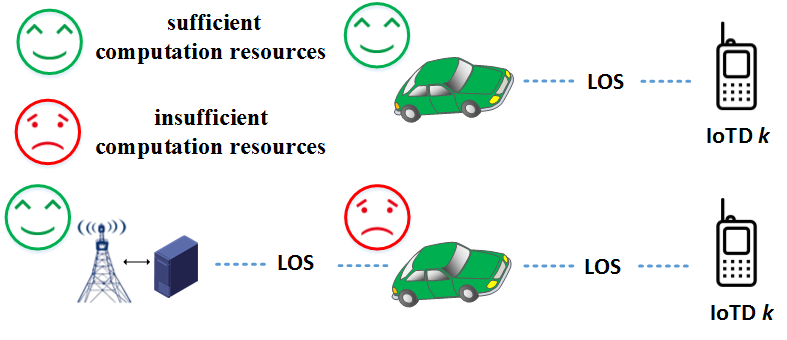}
	\caption{\label{fig2} Two typical scenarios I2V and I2V2S, which are explained in Section \Rmnum{3}-B1 and Section \Rmnum{3}-B2 respectively.}
\end{figure}
\subsection{Communication Model}
\subsubsection{IoTD to Vehicle (I2V)}
For resources available at vehicles on the road, IoTD $k\in \mathcal{K}$ offloads a computation task to one of the passing vehicles. We define $x_k\in \{0, 1\}$ as a offloading decision variable, which indicates whether or not IoTD $k$ offloads their computation bits to the vehicle via a wireless channel (as shown in Fig. 2). Note that, dynamics and randomness of vehicles make the vehicle-assisted VEC extremely complex, and we will focus on solving the characteristics in our future works. Thus, we consider vehicles to be static and do not number them. In other words, we consider that vehicles have the same functions and the same initial resources when computing tasks.
\begin{equation}
	\label{eq1}
	x_k=\left\{
	\begin{aligned}
	1, & ~\textrm{if}~u_k~\textrm{is offloaded}~\textrm{to the vehicle;} \\ 
	0, & ~\textrm{otherwise.}
	\end{aligned}\right. \tag{1}
\end{equation}

Further, it is assumed that the wireless channel between IoTD $k$ and vehicles is based on the line of sight (LOS) link \cite{zhang2019joint,zhou2018computation}, hence the channels between IoTDs and vehicles are modeled by the free space path loss model. Therefore, channel power gain from IoTD $k$ to the vehicle is given by:
\begin{equation}
	\label{eq2}
	h_{kv} = \beta_{0}d_{kv}^{-2}~,\tag{2}
\end{equation}
where $\beta_0$ is channel gain at the reference distance $d_0=1$m, and $d_{kv}$ indicates the distance between IoTD $k$ and the vehicle collecting $u_k$. Similarly, $h_{vm}=\beta_0d_vm^{-2}$ represents channel power gain from the vehicle to RSU $m$.

The total available bandwidth $B$ is equally assigned to each IoTD. Define $p_k$ as transmit power of $u_k$. Specifically, it is assumed that an OFDMA is applied in the system [27]. Thus, transmission rate in bits-per-second (b/s) from IoTD $k$ to the vehicle collecting it for computing is defined as:
\begin{equation}
	\label{eq3}
	R_{k}^{k\rightarrow v} = B_{k}{\log}_{2}\left( {1 + \frac{p_{v}\xi_{0}}{d_{kv}^{2}}} \right),
	\tag{3}
\end{equation}
where $B_k=B/k$ and  $\xi_{0} = \frac{\beta_{0}}{N_{0}B_{k}}$  is expressed as the reference received Signal-to-Noise Ratio (SNR) at IoTD $k$ for $d_0=1$m \cite{zhang2019joint} transmission delay for dispatching a task from IoTD $k$ to the vehicle is given by:
\begin{equation}
	\label{eq4}
	t_{k,c}^{k\rightarrow v} = \frac{\varrho_{k}I_{k}}{R_{k}^{k\rightarrow v}}.
	\tag{4}
\end{equation}

Sometimes, we also face the return of computation results (or instructions). To solve the problem, our scheme considers returning result and optimizes task processing delay. When $u_k$ is completed by the vehicle collecting it, the vehicle delivers its results to the another vehicle passing through IoTD k (Strategy \Rmnum{2} in Fig. 1). Similarly, the wireless channel between the vehicles is based on the LOS link, but the channel $B_k$ is unified as the channel accessed by vheicles participating in $u_k$. Note that, different from previous works \cite{huang2020result,zhang2019joint,ndikumana2019joint,dai2018joint}, We do not ignore results or set results of all tasks to a fixed smaller value for convenience. On the contrary, we integrate the returning results into task offloading by considering output data size to make it more complete. Further, given transmit power $p_v$ of the vehicle, transmission rate in b/s from the transmitting vehicle to the receiving vehicle is defined as:
\begin{equation}
	\label{eq5}
	R_{k}^{v\rightarrow v} = B_{k}{\log}_{2}\left( {1 + \frac{p_{v}\xi_{1}}{d_{vv}^{2}}} \right),
	\tag{5}
\end{equation}
where $d_{vv}$ is the distance from the transmitting vehicle to the receiving vehicle, which is equal to the distance traveled by the vehicle during computing the task, $\xi_{1} = \frac{\beta_{0}}{N_{1}B_{k}}$. In addition, $d_{vv} = \frac{\varrho_{k}{\tilde{c}}_{k}\bar{v}}{F_{V}}$, where $\bar{v}$ denotes the average speed of the vehicle. It can be observed from this formula that the returning time is related to the size of computation bits, so researching on the returning results is necessary. Besides, $d_{vk}$ from the receiving vehicle to IoTD $k$ is equal to $d_{kv}$ from IoTD $k$ to the vehicle collecting $u_k$, so that channel power gain $h_{vk}$ from the vehicle to IoTD $k$ is equal to $h_{kv}$. Thence, transmission rate in b/s is defined as:
\begin{equation}
	\label{eq6}
	R_{k}^{v\rightarrow k} = B_{k}{\log}_{2}\left( {1 + \frac{p_{v}\xi_{1}}{d_{kv}^{2}}} \right).
	\tag{6}
\end{equation}

In summary, transmission delay for returning results of  $u_k$ is given by:
\begin{equation}
	\label{eq7}
	t_{k,c}^{v\rightarrow k} = \frac{\varrho_{k}O_{k}}{R_{k}^{v\rightarrow v}} + \frac{\varrho_{k}O_{k}}{R_{k}^{v\rightarrow k}}\;.
	\tag{7}
\end{equation}

Further, the communication-related delay for $u_k$ is obtained as:
\begin{equation}
	\label{eq8}
	t_{k,c} = t_{k,c}^{k\rightarrow v} + t_{k,c}^{v\rightarrow k} = \frac{\varrho_{k}I_{k}}{R_{k}^{k\rightarrow v}} + \frac{\varrho_{k}O_{k}}{R_{k}^{v\rightarrow v}} + \frac{\varrho_{k}O_{k}}{R_{k}^{v\rightarrow k}}\;.
	\tag{8}
\end{equation}
\setlength{\lineskip}{2.5pt}{}
\setlength{\lineskiplimit}{2.5pt}
For IoTD $k$, energy consumption of uploading data from IoTD $k$ to the vehicle is based on the classic transmission energy model \cite{shu2006joint}, i.e., $E_{k}^{k\rightarrow v} = \left( {p_{k} + p_{k}^{cir}} \right) \cdot t_{k,c}^{k\rightarrow v}$, where $p^{cir}_k$ is circuit power. Following a similar model in \cite{cui2005energy}. $p^{cir}_k$ can be calculated as $p_{k}^{cir} = \vartheta_{k} + ( \frac{1}{\eta} - 1 ) \cdot p_{k}$, where $\vartheta_{k}$ is a transmission-power-independent component that takes possession of the power consumed by circuit, and $\eta$ is power amplifier efficiency. The detailed parameters are summarized in TABLE I. Physically, $\eta$ is defined by the drain efficiency of the RF power amplifier \cite{ren2016dynamic}. Thus, energy consumption of uploading data transmission at IoTD $k$ can be rewritten as:
\begin{equation}
	\label{eq9}
	E_{k}^{k\rightarrow v} = \frac{1}{\eta}p_{k}t_{k,c}^{k\rightarrow v} + \vartheta_{k}t_{k,c}^{k\rightarrow v} = \frac{1}{\eta}\left( {p_{k} + \vartheta_{k,c}} \right)t_{k,c}^{k\rightarrow v}\;,
	\tag{9}
\end{equation}
where $\vartheta_{k,c} = \eta \cdot \vartheta_{k}$ is defined as the equivalent circuit power consumption for data transmission. Energy consumption for receiving data is related to output data size of $u_k$ \cite{ren2015lifetime}. For instance, if IoTD $k$ recevies $O_k$ bits data, energy consumption is $E_{k}^{v\rightarrow k} = \varrho_{k}O_{k}e_{c}$, where $e_c$ is a circuit power for receiving data. Note that, since energy consumption of task transmission and computation is negligible for the vehicle, we do not consider reducing energy consumption by optimizing transmit power of the vehicle.

\begin{table}[!tbp]
	\renewcommand{\arraystretch}{1.3}
	\caption{THE KEY NOTATION}
	\label{table_example}
	\centering
	\begin{tabular}{p {4.8cm}|p {3.2cm}}
		\hline
		\bfseries Notation Definition & \bfseries Notation ( and Value )\\
		\hline
		Set of IoTs & $\left| \mathcal{K} \right| = K = 40$\\
		Set of VEC servers & $\left| \mathcal{M} \right| = M = 6$\\
		Task from IoTD $k$ & $u_k$ \\
		Set of time slots     & $\left| \mathcal{N} \right| = N = 20$ \\
		Size of input data (bits)    & $I_k$ \\
		Size of output data (bits)   & $O_k$ \\
		Required resource for $u_k$  & $\tilde{c}_k$ \\
		Offloading ratio of $u_k$    & $\varrho_{k}$  \\
		Capacitance coefficient of $u_k$  & $k_u=10\,$e-27 \\
		Maximum energy consumption   & $E^{max}_k$  \\
		Maximum permissible latency  & $T^{max}_k$  \\
		Total computation resources  & $F_k,F_V,F_m$  \\
		Computation resources allocated at &  $f_{nm}$ \\
		$u_k$ at RSU $m$  &  \\
		Decision variable   &  $x_k,y_{km}$ \\
		Channel gain at the reference distance  &  $\beta_0=-50dB$ \\
		$d_0=1$ m  & \\
		Channel power gain  & $h_{kv},h_{vm}$ \\
		Maximum transmit power of each  & $p^{max}_k=30$dBm \\
		IoTD for offloading  &  \\
		Maximum transmit power of vehicles  &$p^{max}_v=p^{max}_m=36$dBm \\ 
		and VEC servers for offloading  & \\
		Power amplifier efficiency & $\eta=0.9$ \cite{ren2016dynamic} \\
		Circuit power of IoTDs  & $\vartheta_{k,c}=5$mW \\
		Communication bandwidth  & $B=120$MHz \\
		Energy consumption for receiving   & $e_c=5$nJ/bit  \\
		data  & \\
		Transmission rate in b/s  &  $R^{k\rightarrow v}_k,R^{v\rightarrow v}_k,\cdots$ \\
		Transmission delay   &  $t^{k\rightarrow v}_{k,c},t^{v\rightarrow v}_{k,c},\cdots$  \\
		Excution delay  &   $t_{k,e}^{loc},t_{k,e}^{veh},\cdots$  \\
		Noise power spectrum density  & $N_0=N_1=N_2=$ \\
		 & -130dBm/Hz  \\
		Reference received signal-to-noise   &  $\xi_{0},\xi_{1},\xi_{2}$ \\
		ratio (SNR)  &  \\
		\hline
	\end{tabular}
	\vspace{8pt}
\end{table}
In our model, the total communication-related energy consumption for $u_k$ in a time slot is denoted as:
\begin{equation}
	\label{eq10}
	E_{k}^{com} = \frac{1}{\eta}\left( {p_{k} + \vartheta_{k,c}} \right) \cdot \frac{\varrho_{k}I_{k}}{R_{k}^{k\rightarrow v}} + \varrho_{k}O_{k}e_{c}\,.
	\tag{10}
\end{equation}

Further, the constraint on $p_k$ is defiend as follows:
\begin{equation}
	\label{eq11}
	0 < p_{k} < p_{k}^{max},~\forall k \in \mathcal{K}\,,
	\tag{11}
\end{equation}
where $p_k^{max}$ is the maximum transmit power of IoTD $k$. For convenience, we assume that each IoTD has enough energy in each time slot to transmit their task and receive the related result. In the next research work, we will focus on the random energy harvesting process.

\subsubsection{IoTD to Vehicle to Server (I2V2S)}
When $u_k$ is required to be dispatched to the VEC server, the VNO also determine the server's selection decisions for $u_k$ according to JOET and then informs the vehicle collecting $u_k$ the decision. Moreover, we assume that VEC servers belong to the same VNO, where RAT keeps the track of the available resources of VEC servers. Then, the vehicle offloads the collected task to the target server, and the wireless channel is also based on LOS link.
We denote $y_k^{k\rightarrow m}$ as a decision variable, which indicates whether or not $u_k$ is offloaded from the vehicle collecting it to RSU $m$, which is given by:
\begin{equation}
	\label{eq12}
	y_k^{k\rightarrow m}=\left\{
	\begin{aligned}
		1, & ~\textrm{if}~u_k~\textrm{is offloaded}~\textrm{to RSU}~m \textrm{\,;} \\ 
		0, & ~\textrm{otherwise\,.}
	\end{aligned}\right. \tag{12}
\end{equation}

Similarly, transmission rate $R_k^{v\rightarrow m}$ from the vehicle collecting  $u_k$ to RSU $m$ is given by:
\begin{equation}
	\label{eq13}
	R_{k}^{v\rightarrow m} = B_{k}{\log}_{2}\left( {1 + \frac{p_{v}\xi_{2}}{d_{vm}^{2}}} \right),
	\tag{13}
\end{equation}
where  $\xi_{2} = \frac{\beta_{0}}{N_{2}B_{k}}$, $d_{vm}$ is the distance from the vehicle to RSU $m$. Transmission delay for dispatching a task from IoT $k$ to RSU $m$ is expressed as:
\begin{equation}
	\label{eq14}
	t_{k,c}^{k\rightarrow m} = t_{k,c}^{k\rightarrow v} + t_{k,c}^{v\rightarrow m} = \frac{\varrho_{k}I_{k}}{R_{k}^{k\rightarrow v}} + \frac{\varrho_{k}I_{k}}{R_{k}^{v\rightarrow m}}\,.
	\tag{14}
\end{equation}

Analogously, when RSU $m$ completed $u_k$, RSU $m$ delivers its results to the vehicle passing though IoTD $k$, and then the vehicle delivers the results to IoTD $k$. Given transmit power $p_m$ of RSU $m$, transmission rate from RSU $m$ to the vehicle receiving the results is denoted as:
\begin{equation}
	\label{eq15}
	R_{k}^{m\rightarrow v} = B_{k}{\log}_{2}\left( {1 + \frac{p_{m}\xi_{2}}{d_{mv}^{2}}} \right),
	\tag{15}
\end{equation}
where $d_vm$ is the distance from RSU $m$ the vehicle receiving the results. Transmission delay for returning the results from RSU $m$ to IoT $k$ is given by:
\begin{equation}
	\label{eq16}
	t_{k,c}^{m\rightarrow k} = t_{k,c}^{m\rightarrow v} + t_{k,c}^{v\rightarrow k} = \frac{\varrho_{k}O_{k}}{R_{k}^{m\rightarrow v}} + \frac{\varrho_{k}O_{k}}{R_{k}^{v\rightarrow k}}\,.
	\tag{16}
\end{equation}

Although $u_k$ is transimited to RSU $m$ for offloading through the vehicle, for IoTD $k$, the total communication-related energy consumption is still $E_k^{com}$.

\subsection{Computation Model}
\subsubsection{Computation at Vehicle}
Considering the sensitive delay and the life limited by energy consumption, IoTD $k$ computes $\left(1-\varrho_{k} \right){\tilde{c}}_{k}$ locally, and offloads the rest $ \varrho_{k} \tilde{c}_{k}$ to the vehicle and the VEC server. The initial free resources of all vehicles are given by $F_V$. Since computation resources of vehicles are insufficient, we consider that each vehicle can only compute one task. In other words, the vehicle collecting $u_k$ allocates all resources to $u_k$. Thus, computation delay for $u_k$ at the vehicle is denoted as: 
\begin{equation}
	\label{eq17}
	t_{k,e}^{k\rightarrow v} = \frac{\varrho_{k}{\tilde{c}}_{k}}{F_{V}}\,.
	\tag{17}
\end{equation}

Besides, the local computation time $t_{k,e}^{loc}$ can be given by:
\begin{equation}
	\label{eq18}
	t_{k,e}^{loc} = \frac{\left( {1 - \varrho_{k}} \right){\tilde{c}}_{k}}{F_{k}}\,,
	\tag{18}
\end{equation}
where $F_k$ is the total resources of $u_k$. Referring to Ref. \cite{li2019incentive}, the computation-related energy consumption is expressed as:
\begin{equation}
	\label{eq19}	
	E_{k}^{exe} = k_{u}F_{k}^{2}\left( {1 - \varrho_{k}} \right){\tilde{c}}_{k}\,,\tag{19}
\end{equation}
where $k_u F_k^2$ is the power consumption per CPU cycle and $k_u$ denotes capacitance coefficient that depends on chip architecture \cite{miettinen2010energy}. Thus, the total energy consumption is denoted as $E_{k} = ~E_{k}^{com} + E_{k}^{exe}$. Further, when $u_k$ is offloaded to the vehicle, task processing delay is determined by:
\begin{equation}
	\label{eq20}
	t_{k}^{k\rightarrow v} = {\max\left\{ {t_{k,e}^{loc},~t_{k,c}^{k\rightarrow v} + t_{k,e}^{veh} + t_{k,c}^{v\rightarrow v} + t_{k,c}^{v\rightarrow k}} \right\}}\,.
	\tag{20}
\end{equation}
\subsubsection{Computation at VEC Server}
Sometimes, $u_k$ is dispatched to the VEC server to archieve greater system utility because the vehicle does not have enough computation resources. The VNO derives selection decisions, resource allocation, offloading ratio and transmit power for each IoTD $k$ based on the proposed JOET algorithm. We denote $f_{km}\leq F_m$ as computation resources allocated to $u_k$ at RSU $m$, where $F_m$ as the total computation resources of RSU $m$. While minimizing the task processing delay and the energy consumption of IoTDs, we also considered how to optimize load balancing of servers. Therefore, execution delay for $u_k$ at RSU $m$ is obtained as:
\begin{equation}
	\label{eq21}
	t_{k,e}^{k\rightarrow m} = \frac{\varrho_{k}{\tilde{c}}_{k}}{f_{km}}\,.
	\tag{21}
\end{equation}

For $u_k$ offloaded to RSU $m$, task processing delay is expressed as:
\begin{equation}
	\label{eq22}
	t_{k}^{k\rightarrow m} = {\max\left\{ {t_{k,e}^{loc},~t_{k,c}^{k\rightarrow m} + t_{k,e}^{ser} + t_{k,c}^{m\rightarrow k}} \right\}}\,.
	\tag{22}
\end{equation}

In summary, task processing delay $T_k$ is given by:
\begin{equation}
	\label{eq23}
	T_{k} = x_{k}t_{k}^{k\rightarrow v} + \left( {1 - x_{k}} \right){\sum\limits_{m \in \mathcal{M}}{y_{km}t_{k}^{k\rightarrow m}}}\,.
	\tag{23}
\end{equation}
\subsection{Problem formulation}
According to the above discussion, our objective is to optimize jointly the total energy consumption and task processing delay in consideration of load balancing. We formulate the objective problem as a system utility maximization problem, which is subjected to selection decisions, resource allocation, offloading ratio and transmission power adjustment. Specifically, the utility function is defined as:
\begin{multline}
	\label{eq24}
	\sum_{n\in \mathcal{N}} U[n]\triangleq \sum_{n\in \mathcal{N}} \sum_{k\in \mathcal{K}}\log \left( 1  + \alpha\frac{E_{k}^{max}[n] - E_{k}[n]}{E_{k}^{max}[n]} \notag \right.\\ \left.
	~+ \beta\frac{T_k^{max}[n] - T_k[n]}{T_k^{max}[n]} \right). \tag{24}
\end{multline}
%\end{equation}
where $\alpha$ and $\beta$ are the weight of each items respectively, indicating importance of each item. $E_k^{max}[n]$ and $T_k^{max}[n]$ are the maximum energy consumption and the maximum task processing delay for $u_k$ in time slot $n$. The utility function as a satisfaction function monotonically decreases with $E_k$ and $T_k$. In other words, we minimize $E_k$ and $T_k$ to maximize system utility in each time slot. The logarithmic utility is known as proportional fairness based on \cite{ye2016user,li2008proportional} which is able to achieve load balancing \cite{li2008proportional}. Therefore, we formulate the utility maximization problem as: 
\begin{flalign}
	\label{eq25(1)}	
	~~~~P1:~&\max\limits_{\mathbf{x},\mathbf{y},\mathbf{F},\boldsymbol{\varrho}, \mathbf{P}}U & \notag\\
	&\mbox{s.t.~(1),~(11)~and~(12)\,;} & \notag \\
	&\quad~~ x_{k}t_{k}^{k\rightarrow v} \leq T_{k}^{max},~\forall k \in \mathcal{K}\,;& \tag*{(25a)}
\end{flalign}
\begin{equation}
	\label{eq25(2)}
	\sum_{m \in \mathcal{M}}{y_{km}\left\lbrack x_k t_k^{k\rightarrow v} + 	\left(1-x_k \right)t_k^{k\rightarrow m} \right\rbrack \leq T_k^{max}},~\forall k \in \mathcal{K}\,;
	\tag*{(25b)}
\end{equation}
\begin{flalign}
	\label{eq25(3)}
	&\qquad \qquad \quad \sum_{m \in \mathcal{M}}y_{km} = 1,~\forall k \in \mathcal{K}\,; &\tag*{(25c)} \\
	&\qquad \qquad \quad~ 0 \leq \sum_{k \in \mathcal{K}}{y_{km}f_{km}} \leq F_{m},~\forall m \in \mathcal{M}\,; & \tag*{(25d)} \\
	&\qquad \qquad \quad~ 0 \leq {\sum\limits_{m \in \mathcal{M}}{y_{km}\varrho_{k}}} \leq 1,~\forall k \in \mathcal{K}\,; & \tag*{(25e)} \\
	&\qquad \qquad \quad~ 0 \leq \varrho_{k} \leq 1,~\forall k \in \mathcal{K}\,, & \tag*{(25f)}
\end{flalign}
where constraints (25a) and (25b) guarantee that task processing delay cannot exceed the maximum permission delay $T_k^{max}$, and constraint (25c) are used to ensure that $u_k$ can only be computed by one server. Constraint (25d) means that the sum of allocated resources to tasks which select RSU $m$ does not exceed $F_m$. Specifically, the lower bound of $p_k$ is also constrained in constraints (25a) and (25b). Constraints (25e) and (25f) denote offloading ratio of $u_k$ cannot exceed to 1.  In addition, as mentioned earlier, $u_k$'s allocated resources are $F_V<f_{km}$ when  $u_k$ is offloaded to the vehicle. This is because $f_{km}$ needs to make up for transmission time from the vehicle to the server.

Our objective is to optimize jointly the total energy consumption and task processing delay under the premise of ensuring that the task is completed. Similar to Ref. \cite{dai2018joint}, due to the integer constraint $y_{km}$, $P1$ is a mixed integer nonlinear programming problem. But the problem to be solved in this paper is more difficult than Ref. \cite{dai2018joint} because we not only need to optimize task processing delay, but also the total energy consumption. In the following section, we will pour attention to how to solve this more intractable problem.
\section{PROBLEM DECOMPOSITION AND SOLUTION}
Since there are highly complex coupling among optimization variables and mixed combinatorial feature in $P1$, it is quite challenging to solve $P1$. Let a binary matrix , a binary matrix $\mathbf{x} = \left\{ x_{k} \middle| \forall k \in \mathcal{K} \right\}$, a binary matrix $\mathbf{y} = \left\{ y_{km} \middle| \forall k \in \mathcal{K},~m \in \mathcal{M} \right\}$, a matrix $\mathbf{F} = \left\{ f_{km} \middle| \forall k \in \mathcal{K},~m \in \mathcal{M} \right\}$, a matrix $\boldsymbol{\varrho} = \left\{ \varrho_{k} \middle| \forall k \in \mathcal{K} \right\}$ and a matrix $\mathbf{P} = \left\{ p_{k} \middle| \forall k \in \mathcal{K} \right\}$, denote offloading decision, server selection decision, resource allocation, offloading ratio and transmit power, respectively. In this section, we aim to decouple $\mathbf{x}$ and $\mathbf{y}$, $\mathbf{F}$, $\boldsymbol{\varrho}$ and $\mathbf{P}$ into two subproblems (i.e., optimization of selection decisions, optimization of resource allocation, offloading ratio and transmit power). That is, we firstly determine $\mathbf{x}$ and $\mathbf{y}$ under given $\mathbf{F}$, $\boldsymbol{\varrho}$ and $\mathbf{P}$, and then $\mathbf{F}$, $\boldsymbol{\varrho}$ and $\mathbf{P}$ under obtained $\mathbf{x}$ and $\mathbf{y}$, and repeat this process until convergence.

To facilitate the introduction of JOET, we define the follwing intermediate variables:
\begin{equation}
	\label{eq26}
	\left\{
	\begin{aligned}
		a_{km}=\frac{E_{k}^{max} - E_{k}^{com}}{E_{k}^{max}}\,; \\ 
		b_{km}=\frac{T_{k}^{max} - T_{k}^{pro}}{T_{k}^{max}}\,.
	\end{aligned}\right. \tag{26}
\end{equation}

Therefore, taking time slot $n$ as an example, the utility function can be rewritten as:
\begin{equation}
	\label{eq27}
	U\left\lbrack n \right\rbrack \triangleq {\sum\limits_{k \in \mathcal{K}}{\log\left( {1 + \alpha a_{km} + \beta b_{km}} \right)}}\,. \tag{27}	
\end{equation}
\vspace{-3mm}
\subsection{Optimization of Selection Decisions}
The selection decisions problem under given $\mathbf{F}$, $\boldsymbol{\varrho}$ and $\mathbf{P}$ can be formulated as: 
\begin{flalign}
	\label{eq28}	
	~~~~P1.1:~&\max\limits_{\mathbf{x},\mathbf{y}}U & \notag\\ 
	&\mbox{s.t.~(1),~(12),~(25a),~(25b)~and~(25c)}\,. & \tag{28}
\end{flalign}

Since all the indicator variables $\mathbf{x}$ and $\mathbf{y}$ are binary and the objective function is nonlinear with respect to $\mathbf{x}$ and $\mathbf{y}$, $P1.1$ is also a MINLP problem \cite{dai2018joint,kleinberg2006algorithm}. Thus, we cannot obtain the optimal $\mathbf{x}$ and $\mathbf{y}$ in polynomial time. To solve it with low complexity, we employ an approximation algorithm, which firstly relaxes $P1.1$ as a continuous nonlinear programming problem and uses the Block Successive Upper-bound Minimization (BSUM) method to solve it.

To better describe optimization of selection decisions, we briefly introduce BSUM. The advantages of BSUM reside in both solution speed and problem decomposability \cite{ndikumana2019joint,hong2015unified}. The standard form of BSUM is:
\begin{flalign}
	\label{eq29}
	~~\min_{\mathbf{z}}&~g(\mathbf{z}_1,\mathbf{z}_2,\ldots,\mathbf{z}_J), &\notag \\
	&\mbox{s.t.~} \mathbf{z}_j\in \mathcal{Z}_j, \forall j\in \mathcal{J},~j=1,\ldots,J\,, &\tag{29}
\end{flalign}
where $\mathcal{Z}:=\mathcal{Z}_{1} \times \cdots \times \mathcal{Z}_{J}$, $g(\cdot )$ is a continuous function and  $I$ is the set of indexes. For $j=1,\;\ldots\,,\,J$, we consider $\mathcal{Z}_j$ as a closed convex set and $z_j$ as a block of variables. Similar to Block Coordinate Descent (BCD), at each iteration $i$, a single block of variables is optimized by solving the following problem:
\begin{equation}
	\label{eq30}
	\mathbf{z}_{j}^{i} \in {\underset{\mathbf{z}_{j} \in \mathcal{Z}_{j}}{\textrm{argmin}}{~g\left( {\mathbf{z}_{j},\mathbf{z}_{- j}^{i - 1}} \right)}}\,, \tag{30}
\end{equation}
where $\mathbf{z}_{- j}^{i - 1}:=( {\mathbf{z}_{1}^{i - 1},\ldots,\mathbf{z}_{j - 1}^{i - 1},\mathbf{z}_{j + 1}^{i - 1},\,\ldots\,,\,\mathbf{z}_{j}^{i - 1}} ),~\mathbf{z}_{l}^{i} = \mathbf{z}_{l}^{i - 1}$ for $j\neq l$. Since $\mathbf{x}$ and $\mathbf{y}$ are integer variables, $P1.1$ is a non-convex function to make the problem more difficult, and BCD does not always guarantee convergence. BSUM is an improved method of BCD. More details are presented in Ref. \cite{ndikumana2019joint} and Ref. \cite{hong2015unified}. 

Thus, we consider BSUM as a suitable candidate method for solving it by focusing on solving perblock problems. Using BSUM to solve $P1.1$, we need two steps:

i) Step 1: we introduce a proximal function which is a convex optimization problem and an upper bound of $P1.1$ by adding quadratic penalization \cite{ndikumana2019joint};

ii) Step 2: By negating the system utility function to turn it into a minimization problem as $B(x,y):=-U$.  Then we minimize the proximal upper-bound function and ensure that the upper-bound function takes steps proportional to the negative value of gradient.

At each iteration $i$, $\forall j\in \mathcal{J}$, we define the proximal upper-bound function $B_j$, which is convex and the proximal upper-bound of the objective function defined in (28). To make $B_j$ convex, we  introduction a quadratic penalization, as follows:

\begin{align}
	\label{eq31}
	B_{j}\left( {\mathbf{z}_{j},\mathbf{x}^{(i)},\mathbf{y}^{(i)}} \right) := & B\left( {\mathbf{z}_{j},\mathbf{x}^{({i - 1})},\mathbf{y}^{({i - 1})}} \right) \notag\\
	& + \frac{\delta_{j}}{2}\|  {\mathbf{z}_{j} - \mathbf{x}^{(i - 1)}} \|^{2}\,. \tag{31}
\end{align}

Herein, (31) is the proximal upper-bound of (28) and it also can be employed to other vector of variable $\mathbf{y}$. where $\delta_{j}>0$ indicates the positive penalty parameter. Further, eq. (31) is a convex optimization problem because of its quadratic term $\frac{\delta_{j}}{2}\left\|  {\mathbf{z}_{j} - \mathbf{x}^{(i - 1)}} \right\|^{2}$, where $\mathbf{x}^{(i-1)}$ and $\mathbf{y}^{(i-1)}$ are considered to be the solution of the previous step $(i-1)$. Next, we update the solution by the following formula:
\begin{align}
	\label{eq3233}
	\mathbf{x}^{(i + 1)} \in {\min\limits_{\mathbf{x}}{B_{j}\left( {x^{(i + 1)},\mathbf{x}^{(i)},\mathbf{y}^{(i)}} \right)}}\,, \tag{32} \\
	\mathbf{y}^{(i + 1)} \in {\min\limits_{\mathbf{y}}{B_{j}\left( {y^{(i + 1)},\mathbf{x}^{(i)},\mathbf{y}^{(i)}} \right)}}\,.  \tag{33}
\end{align}

We adopt the BSUM method of Ref. \cite{hong2015unified} to solve (32) and (33). The solution process is presented in the following Algorithm 1 (BSUM for Optimization of Selection Decisions), which is consistent with BSUM in Ref. \cite{hong2015unified}. Herein, we employ the Cyclic rule to select each coordinate $j\in \mathcal{J}$ \cite{hong2015unified}. This is because a task is transmitted to the vehicle first, and only when the vehicle cannot meet the maximum delay of the task or achieve a lower system benefit, the task will be offloaded to the server. \vspace{1mm}
\begin{algorithm}[h]
	\caption{BSUM for Optimization of Selection Decisions} %算法的名字
	%\hspace*{0.02in} {\bf Input:} %算法的输入， \hspace*{0.02in}用来控制位置，同时利用 \\ 进行换行
	\textbf{Input:} $\mathbf{x}$ and $\mathbf{y}$ \\
	% \hspace*{0.02in} {\bf Output:} %算法的结果输出
	\textbf{Output:} $\mathbf{x}^*$ and $\mathbf{y}^*$
	\begin{algorithmic}[1]
		\State Initialize $i=0,\epsilon>0$; % \State 后写一般语句
		\State Find initial feasible points $(mathbf{x}^{(0)},\mathbf{y}^{(0)})$
		\Repeat
		\State Choose index set I by Cyclic rule, i.e., 1,2,1,$\ldots$;
		\State Relax $\mathbf{x}$, and let $\mathbf{x}^{(i+1)}\in \min\limits_{\mathbf{x}}{B_{j}\left( {x^{(i + 1)},\mathbf{x}^{(i)},\mathbf{y}^{(i)}} \right)}$;
		\State Set $x_{k}^{({i + 1})} = x_{k}^{(i)},~\forall k \notin \mathcal{J}$;
		\If {$x_{k}^{(i)} \geq \rho$}
		\State $x_{k}^{({i + 1})} = 1$;
		\Else
		\State $x_{k}^{({i + 1})} = 0$;
		\EndIf \label{endif}
		\State Jump to step 4, obtain $y^{(i+1)}$ by solving (33);
		\State $i=i+1$
		\Until $\| \frac{B_{j}^{(i)} - B_{j}^{(i + 1)}}{B_{j}^{(i)}} \|\leq \epsilon$
		\State Then, consider $\mathbf{x}^{*} = \mathbf{x}^{(i + 1)}$, $\mathbf{y}^{*} = \mathbf{y}^{(i + 1)}$;
	\end{algorithmic}
\end{algorithm}

Algorithm 1 (BSUM for Optimization of Selection Decisions) can be considered as a generalized form of BCD, which optimizes the upper-bound function of the original objective function block by block. In Algorithm 1, we determine the offloading decision $x_k$ for each $u_k$ by first relaxing $x_k$ and then (32). Note that, relaxing details will be introduced below, and we first introduce the complete structure of Algorithm 1. If $x_k=1$, the user transmit its task to a passing vehicle; else, $u_k$ will be computed locally. Similar to $x$,  we further determine the server selection $y_{km}$ for  each $u_k$. If $y_{km}=1$, then  $u_k$ will be offload to RSU m; else, $u_k$ will be computed the vehicle. At each iteration $i+1$, we update the solution by solving (32) and (33) until $\| \frac{B_{j}^{(i)} - B_{j}^{(i + 1)}}{B_{j}^{(i)}} \|\leq \epsilon$, where $\epsilon\in (0,1)$. 

When Algorithm 1 is used to solve (32) and (33) , we first relax the vectors of variables $\mathbf{x}$ and $\mathbf{y}$ into a continuous value in [0,1], and then we use a threshold rounding method \cite{shmoys1993approximation} to compel the relaxed x and y to be vectors of binary variables.

Taking $\mathbf{x}$ as an example, we set the feasible solution $x_k^{i*}\in \mathbf{x}^i$ as follows:
\begin{equation}
	\label{eq34}
	x_k^{i*}=\left\{
	\begin{aligned}
		1, & ~\textrm{if}~x_k^{i*}\geq \rho\,;\\ 
		0, & ~\textrm{otherwise\,.}
	\end{aligned}\right. \tag{34}
\end{equation}
where $\rho \in (0,1)$ is a positive rounding threshold. The above round method can also be used to  $\mathbf{y}$. However, the obtained binary solution may break communication and computational constraints (25a) and (25b). To avoid this issue after rounding, we solve (31) in the form of $B_{j} + \varepsilon\Delta$, where $\varepsilon$ is a constant weight. Then, constraints are modified as follows:
\begin{align}
	\label{eq3536}
	x_{k}t_{k}^{k\rightarrow v} \leq T_{k}^{max} + \mathrm{\Delta}_{x},~\forall k \in \mathcal{K}\,; \tag{35} \\
	{\sum\limits_{m \in \mathcal{M}}{y_{km}t}_{k}^{k\rightarrow m}} \leq T_{k}^{max} + \mathrm{\Delta}_{y},~\forall k \in \mathcal{K}\,, \tag{36}
\end{align}
where $\mathrm{\Delta}_{x}$ and $\mathrm{\Delta}_{y}$ are the maximum break of communication and computation resources, $\Delta = \mathrm{\Delta}_{x} + \mathrm{\Delta}_{y}$. Unlike Ref. \cite{ndikumana2019joint}, we integrate communuication and computation resources into task processing time to minimize the time. $\mathrm{\Delta}_{x}$ and $\mathrm{\Delta}_{y}$ are given by:
\begin{align}
	\label{eq3738}
	\mathrm{\Delta}_{x} = {\max\left\{ 0,~x_{k}t_{k}^{k\rightarrow v} - T_{k}^{max} \right\}},~\forall k \in \mathcal{K}\,; \tag{37} \\
	\mathrm{\Delta}_{y} = {\max\{ 0,{\sum\limits_{m \in \mathcal{M}}{y_{km}t}_{k}^{k\rightarrow m}} - T_{k}^{max} \}},~\forall k \in \mathcal{K}\,. \tag{38}
\end{align}

If $\mathrm{\Delta}_{x}$ and $\mathrm{\Delta}_{y}$, we didn't break constraints (25a) and (25b). Next, we take the integrality gap to measure the ratio between the feasible solutions of $B_j$ and $B_{j} + \varepsilon\Delta$. According to definition and proof of integrality gap in Ref. \cite{hong2015unified}, we can copy the following definition:

\hangindent 1em
\begin{myDef}[\textbf{Integrality gap}]
	~Given problem (31) and its rounded problem $B_{j} + \varepsilon\Delta$,  the integrality gap is given by:
	\begin{equation}
		\label{eq39}
		\vartheta = {\min\limits_{\mathbf{x},\mathbf{y}}\frac{B_{j}}{B_{j} + \varepsilon\Delta}}\,. \tag{39}
	\end{equation}
	We obtain $B_j$ through relaxation of variables $\mathbf{x}$ and $\mathbf{y}$ while we get $B_{j} + \varepsilon\Delta$ after rounding the relaxed $\mathbf{x}^{(i)}$ and $\mathbf{y}^{(i)}$. Then, when $\vartheta\left( \vartheta \leq 1 \right)$ is close to 1 \cite{hong2015unified}, we achieve the best rounding. That is if $\vartheta = 1$, then $\mathrm{\Delta}_{x}=0$ and $\mathrm{\Delta}_{y}=0$.
\end{myDef}

Further, according to Ref. \cite{hong2015unified} and Ref. \cite{ndikumana2019joint}, we have the following remark:
\begin{myRem}[\textbf{Convergence}]
	~BSUM for optimization of selection decisions takes $\mathcal{O}\left( {\log\left( 1/\epsilon \right)} \right)$ to converge to an $\epsilon$-optimal solution.
\end{myRem}
According to Section \Rmnum{2}-D  of Ref. \cite{hong2015unified}, BSUM takes at most $c/\epsilon$ iterations to find an $\epsilon$-optimal solution, where $c>0$ is a constant only related to the description of a problem. Further, for different special forms of BSUM, the constant $c$ in front of $1/\epsilon$ can be significantly refined so that it is independent of problem dimension \cite{hong2015unified,zhan2020mobility}. So BSUM takes  $\mathcal{O}\log(1/\epsilon)$ to coverage to an  $\epsilon$-optimal solution.
\subsection{Joint Optimization of Resource Allocation,  Offloading Ratio and Transmit Power}
Given selection decisions combination vectors $\mathbf{z}$, $P1$ is a joint optimization of resource allocation, offloading ratio and transmit power problem, which is formulated as:
\subsubsection{Computation at Vehicle}
Unlike the server, since the vehicle can only compute one task at a time and do its best to complete it, we just need to optimize offloading ratio and transmit power of $u_k$. However, according to (20), $t_k^{k\rightarrow v}$ is nondifferentiable with respect to $\varrho_{k}$ and $p_k$. To solve this problem, we do the following approximate:
\begin{flalign}
	\label{eq40}	
	~~P1.2:&\max\limits_{\mathbf{F},\boldsymbol{\varrho},\mathbf{P}}U \notag\\
	&\mbox{s.t.~(11),~(25a),~(25b),~(25d),~(25e)~\textrm{and}~(25f)}\,. \tag{40}
\end{flalign}

Since the resource allocation constraints on the vehicle and the server are different, we solve them separately for above two case.
\setlength{\lineskip}{2.5pt}
\setlength{\lineskiplimit}{2.5pt}
\begin{align}
	\label{eq41}
	t_{k}^{k\rightarrow v} & \leq \frac{\left( {1 - \varrho_{k}} \right){\tilde{c}}_{k}}{F_{k}} + \frac{\varrho_{k}I_{k}}{R_{k}^{k\rightarrow v}} + \frac{\varrho_{k}{\tilde{c}}_{k}}{F_{V}} + \frac{\varrho_{k}O_{k}}{R_{k}^{v\rightarrow v}} + \frac{{\varrho_{k}O}_{k}}{R_{k}^{v\rightarrow k}}  \notag \\
	& = \varrho_{k}r_{kv} + \frac{\varrho_{k}O_{k}}{R_{k}^{v\rightarrow v}}~ + \frac{{\tilde{c}}_{k}}{F_{k}}\,, \tag{41}
\end{align}
where $r_{kv} = \frac{I_{k}}{R_{k}^{k\rightarrow v}} + \frac{\left( {F_{k} - F_{V}} \right){\tilde{c}}_{k}}{F_{k}F_{V}} + \frac{O_{k}}{R_{k}^{v\rightarrow k}}$. From (35), we can obtain that the upper-bound of $t_{k}^{k\rightarrow v}$ is $\varrho_{k}r_{kv} + O_{k}/R_{k}^{v\rightarrow v} + {\tilde{c}}_{k}/F_{k}$. Further, we first rewrite $P1.2$ under the worst case delay. Thus, $P1.2$ is transformed into $P1.2.1$.
\begin{flalign}
	\label{eq42}	
	~~~~P1.2.1:~&\max\limits_{\boldsymbol{\varrho},\mathbf{P}}U \notag\\
	&\mbox{s.t.~(11),~(25a),~(25e)~and~(25f)}\,. \tag{42}
\end{flalign}
Furthermore, according to Lemma 1, $P1.2.1$ is a nonconvex optimization problem. Moreover, offloading ratio and transmit power are coupled with each other in both the objective function and constraints. Thus, we have developed a low-complexity algorithm by decoupling two variables. In other words, we derive $\boldsymbol{\varrho}$ under given $\mathbf{P}$ and then derive $\mathbf{P}$ under given $\boldsymbol{\varrho}$ and repeat this process until convergence.
\begin{myLem}
	$P1.2.1$ is a nonconvex problem.
	\begin{myPro}
		~See the Appendix A. 

		$\hfill \qedsymbol$ 
	\end{myPro}
\end{myLem}
\vspace{-1.5mm}
\textit{(1) Offloading Ratio}: Under given $\mathbf{P}$, we can rewritten $P1.2.1$ as the offloading ratio adjustment problem as follows:
\begin{flalign}
	\label{eq43}	
	~~~~P1.2.1.1:~&\max\limits_{\boldsymbol{\varrho}} \log\left( {1 + \alpha a_{k0} + \beta\frac{T_{k}^{max} - t_{k}^{k\rightarrow v}}{T_{k}^{max}}} \right) & \notag\\
	&\mbox{s.t.~(25a),~(25e)~and~(25f)}\,. & \tag{43}
\end{flalign}

Since ${{\partial^{2}U}/{\partial\boldsymbol{\varrho}^{2}}} \leq 0$ based on Lemma 1, $P1.2.1.1$ is convex combining with linear constraint $\varrho_k$. In addition, under given $\mathbf{F}$ and $\mathbf{P}$, $\boldsymbol{\varrho}$ are independent of each other. Thus, we apply primal-dual Lagrangian method to solve this problem. The Lagrangian function is constructed as follows:
\begin{flalign}
	&\mathcal{L}\left( {\boldsymbol{\varrho},\boldsymbol{\zeta},\boldsymbol{\varsigma},\boldsymbol{\tau}} \right) = {\sum_{k \in \mathcal{K}}{\log\left( {1 + \alpha a_{k0} + \beta\frac{T_{k}^{max} - t_{k}^{k\rightarrow v}}{T_{k}^{max}}} \right)}} - &\notag \\
	&\zeta_{k}\left( {\varrho_{k}r_{kv} + \frac{\varrho_{k}O_{k}}{R_{k}^{v\rightarrow v}} + \frac{{\tilde{c}}_{k}}{F_{k}} - T_{k}^{max}} \right) - \varsigma_{k}\left( {\varrho_{k} - 1} \right) + \tau_{k}\varrho_{k}\,, & \tag{44}
\end{flalign}
where $\zeta_{k}, \varsigma_{k}$ and $\tau_{k}$ are the Lagrangian multipliers. Since $\varrho_k$ is independent, the following complementary Slackness conditions must be satisfied based on the KKT conditions \cite{boyd2004convex}:
\begin{flalign*}
	&& \zeta_{k}\left( {\varrho_{k}r_{kv} + \frac{\varrho_{k}O_{k}}{R_{k}^{v\rightarrow v}} + \frac{{\tilde{c}}_{k}}{F_{k}} - T_{k}^{max}} \right) = 0; &\qquad \notag \\
	&& 0 \leq \varrho_{k} \leq 1; &\qquad \notag \\
	&& \varsigma_{k}\left( {\varrho_{k} - 1} \right) = 0; &\qquad \notag \\
	&& \tau_{k}\varrho_{k} = 0;  &\qquad \notag \\
	&& \gamma_{k} > 0,\phi_{k} \geq 0,\psi_{k} \geq 0.&\qquad \tag{45}
\end{flalign*}

Further, the first derivative of $\varrho_k$ can be obtained as follows:
\setlength{\lineskip}{2.5pt}
\setlength{\lineskiplimit}{2.5pt}
\begin{flalign}
	\frac{\partial\mathcal{L}\left( {\boldsymbol{\varrho},\boldsymbol{\zeta},\boldsymbol{\varsigma},\boldsymbol{\tau}} \right)}{\partial\boldsymbol{\varrho}} =& - \frac{1}{\ln\left( 2 \right)}\frac{I_{k}\omega_{1}/R_{k}^{k\rightarrow v} + \omega_{2} + \omega_{3}}{\omega_{0}} \notag \\
	&- ~\zeta_{k}\left( r_{kv} + \omega_{3} \right) - \varsigma_{k} + \tau_{k}\,, \tag{46} 
\end{flalign}
where $\omega_{0} = 1 + \alpha a_{k0} + \beta b_{k0}, \omega_{1} = \frac{\alpha\left( {p_{k} + \vartheta_{k,c}} \right)}{\eta E_{k}^{max}} + \frac{\beta}{T_{k}^{max}}, \omega_{2} = \frac{\beta}{T_{k}^{max}}\left( {\frac{\left( {F_{k} - F_{V}} \right){\tilde{c}}_{k}}{F_{k}F_{V}} + \frac{O_{k}}{R_{k}^{v\rightarrow k}}} \right) + \frac{\alpha\left( O_{k}e_{c} - k_{u}F_{k}^{2}{\tilde{c}}_{k} \right)}{E_{k}^{max}}$ and $\omega_{3} = \frac{O_{k}}{R_{k}^{v\rightarrow v}} + \frac{\varrho_{k}{\tilde{c}}_{k}\bar{v}}{{d_{vv}^{3}F}_{V}} \times \frac{2{B_{k}O}_{k}p_{v}\xi_{1}}{{\ln\left( 2 \right)}\left( R_{k}^{v\rightarrow v})^{2}\left( {1 + \frac{p_{v}\xi_{1}}{d_{vv}^{2}}} \right) \right.}$. 

By taking $\partial\mathcal{L}\left( {\boldsymbol{\varrho},\boldsymbol{\zeta},\boldsymbol{\varsigma},\boldsymbol{\tau}} \right)/\partial\boldsymbol{\varrho} = 0$, $\boldsymbol{\varrho}^{*}$ can be obtained. However, considering the slackness condition of $\varsigma_{k}\left( {\varrho_{k} - 1} \right) = 0$ and $\tau_{k}\varrho_{k} = 0$, we face the following four possible cases: 

Case 1: $\varsigma_{k} > 0$ and $\tau_{k} = 0$. Then, $\varrho_{k} = 1$. It means that the whole computation bits of $u_k$ is offloaded to the vehicle;

Case 2: $\varsigma_{k} = 0$ and $\tau_{k} > 0$. Then, $\varrho_{k} = 0$. It means that $u_k$ is computed locally.

Case 3: $\varsigma_{k} > 0$ and $\tau_{k} > 0$. In the case, there is not $\varrho_k$ satisfing all KTT conditions.

Case 4: $\varsigma_{k} = 0$ and $\tau_{k} = 0$. Then, $\zeta_{k}( ~\varrho_{k}r_{kv} + O_{k}/R_{k}^{v\rightarrow v} + {\tilde{c}}_{k}/F_{k} - T_{k}^{max}) = 0$ and $\partial\mathcal{L}\left( {\boldsymbol{\varrho},\boldsymbol{\zeta},\boldsymbol{\varsigma},\boldsymbol{\tau}} \right)/\partial\boldsymbol{\varrho} = 0$. Due to $d_{vv} = \varrho_{k}{\tilde{c}}_{k}\bar{v}/F_{V}$, the approximate optimal  $\varrho^*$ can be obtained by iteration methods of nonlinear equations \cite{awwal2019modified}. Therefore, 
\begin{flalign}
	\label{eq47}
	\qquad &\zeta_{k} = - \frac{1}{\ln\left( 2 \right)}\frac{I_{k}\omega_{1}/R_{k}^{k\rightarrow v} + \omega_{2} + \omega_{3}}{\omega_{0}\left( {r_{kv} + \omega_{3}} \right)}\,; \notag \\
	& \varrho_{k} = \varrho_{k}^{j*}, \tag{47}
\end{flalign}
where $\varrho_{k}^{j*}$ and denotes that $\varrho_k$ reaches convergence at the $j$th iteration. According to above analysis, we obtain the solution $\varrho_k$ as: 
\begin{equation}
	\label{eq48}
	\varrho_k= \left\{
	\begin{aligned}
		&\varrho_{k}^{j*},~&\textrm{if}~\varsigma_{k} = 0~\textrm{and}~\tau_{k} = 0\,; \\
		&1,~&\textrm{if}~\varsigma_{k} > 0~\textrm{and}~\tau_{k} = 0\,; \\
		&0,~&\textrm{if}~\varsigma_{k} = 0~\textrm{and}~\tau_{k} > 0\,. 
	\end{aligned}\right. \tag{48}
\end{equation}

\textit{(2) Transmit Power: } Further, $P1.2.1$ is transformed into $P1.2.1.2$ under given $\boldsymbol{\varrho}$.
\begin{flalign}
	\label{eq49}	
	~~~~P1.2.1.1:~&\max\limits_{\mathbf{P}} \log( {1 + \alpha a_{k0} + \beta b_{k0}}) & \notag\\
	&\mbox{s.t.~(11)~and~(25a)}\,. & \tag{49}
\end{flalign}

Similarly, ${{\partial^{2}U}/{\partial\mathbf{P}^{2}}} \leq 0$ based on \textit{Lemma 1}. Combining with the linear constraint $p_k$, $P1.2.1.2$ is also a convex optimization problem. The process of solving this problem is the same as that of $P1.2.1.1$. Due to limited space of this paper, we will not repeat it here. Referring to $P1.2.1.1$, the optimal solution of $p_k$ is obtained as:
\setlength{\lineskip}{2.5pt}
\setlength{\lineskiplimit}{2.5pt}
\begin{equation}
	\label{eq50}
	p_k^*= \left\{
	\begin{aligned}
		&\frac{d_{kv}^{2}}{\xi_{0}}( {2^{\frac{\varrho_{k}I_{k}}{\Omega_{k}B_{0}}} - 1} ),~&\textrm{if}~\phi_{k} = 0~\textrm{and}~\psi_{k} = 0\,;\\
		&1,~&\textrm{if}~\phi_{k} > 0~\textrm{and}~\psi_{k} = 0\,,
	\end{aligned}\right. \tag{50}
\end{equation}
where $\Omega_{k} = T_{k}^{max} - \frac{\varrho_{k}O_{k}}{R_{k}^{v\rightarrow v}} - \frac{{\tilde{c}}_{k}}{F_{k}} - \frac{\varrho_{k}O_{k}}{R_{k}^{v\rightarrow k}} - \frac{\left( F_{k} - F_{V} \right)\varrho_{k}{\tilde{c}}_{k}}{F_{k}F_{V}}$.
\subsubsection{Computation at VEC Server}
On the contrary, when the task is offloaded to the server, we not only need to sustainably reduce energy consumption and task processing delay of $u_k$, but also need to balance the load of servers by optimizing the allocation of computation resources. Similarly, according to (22), $t_k^{k\rightarrow m}$ is nondifferentiable with respect to $\varrho_k$ and $p_k$. Thus, we agian approximate $t_k^{k\rightarrow m}$ as follows:
\begin{align}
	\label{eq51}
	t_{k}^{k\rightarrow m} \leq & \frac{\left( {1 - \varrho_{k}} \right){\tilde{c}}_{k}}{F_{k}} + \frac{\varrho_{k}I_{k}}{R_{k}^{k\rightarrow v}} + \frac{\varrho_{k}I_{k}}{R_{k}^{v\rightarrow m}} + \frac{\varrho_{k}{\tilde{c}}_{k}}{f_{km}} + \frac{\varrho_{k}O_{k}}{R_{k}^{m\rightarrow v}} \notag \\
	&+ \frac{\varrho_{k}O_{k}}{R_{k}^{v\rightarrow k}} = \varrho_{k}r_{km} + \frac{{\tilde{c}}_{k}}{F_{k}}\,, \tag{51}
\end{align}
where $r_{km} = \frac{I_{k}}{R_{k}^{k\rightarrow v}} + \frac{I_{k}}{R_{k}^{v\rightarrow m}} + \frac{\left( {F_{k} - f_{km}} \right){\tilde{c}}_{k}}{F_{k}f_{km}} + \frac{O_{k}}{R_{k}^{m\rightarrow v}} + \frac{O_{k}}{R_{k}^{v\rightarrow k}}$. Similarly, $\varrho_{k}r_{km} + {\tilde{c}}_{k}/F_{k}$ is the upper bound of $t_k^{k\rightarrow m}$. $P1.2$ is transformed into $P1.2.2$ under the worst case delay.
\begin{flalign}
	\label{eq52}	
	~~~~P1.2.2:~&\max\limits_{\mathbf{F},\boldsymbol{\varrho},\mathbf{P}} U & \notag\\
	&\mbox{s.t.~(11),~(25b),~(25d),~(25e)~and~(25f)}\,. & \tag{52}
\end{flalign}
\begin{myLem}
	$P1.2.2$ is a nonconvex problem.
	\begin{myPro}
		~See Appendix B on the appendix file. 
		
		$\hfill \qedsymbol$ 
	\end{myPro}
\end{myLem}
\vspace{-1.5mm}
According to Lemma 2, $P1.2.2$ is also a nonconvex problem. In addition, $\mathbf{F}$,$\boldsymbol{\varrho}$ and $\mathbf{P}$ are highly coupled with each other. Therefore, we further decouple three variables to develop a low-complexity method. Specifically, we derive $\mathbf{F}$ under given $\boldsymbol{\varrho}$ and $\mathbf{P}$ and then derive $\boldsymbol{\varrho}$ and $\mathbf{P}$ under given $\mathbf{F}$, and repeat this process until convergence.

\textit{(1) Computation Resources:} Under given $\boldsymbol{\varrho}$ and $\mathbf{P}$, we can rewritten $P1.2.2$ as the computation resource allocation problem $P1.2.2.1$ as:
\begin{flalign}
	\label{eq53}	
	~~~~P1.2.2.1:~&\max\limits_{\mathbf{F}} U & \notag\\
	&\mbox{s.t.~(25b)~and~(25d)}\,. & \tag{53}
\end{flalign}

Since $f_{km}$ is a linear constraint, P1.2.2.1 is a convex optimization problem as ${{\partial^{2}U}/{\partial\mathbf{F}^{2}}} \leq 0$. However, for tasks who are offloaded to RSU $m$. $\mathbf{F}$ are highly coupled with each other. In other words, if the resource allocation of one of the tasks is increased, the resource allocation of other tasks will decrease. Therefore, we construct Lagrangian function and iteratively approach a optimal solution.
\begin{flalign}
	\label{eq54}
	&\mathcal{L}\left( {\mathbf{F},\boldsymbol{\lambda},\boldsymbol{\mu},\boldsymbol{\theta}} \right) = {\sum_{k\in \mathcal{K}}{\log\left( 1 + \alpha a_{km} + \beta b_{km} \right)}}- {\sum_{k\in \mathcal{K}}\lambda_{k}} & \notag \\
	&\times ( {\sum_{m\in \mathcal{M}}{y_{km}t_{km} - T_{k}^{max}}} ) + {\sum_{k\in \mathcal{K}} {\sum_{m\in \mathcal{M}}{\mu_{km}( y_{km}f_{km} }} - F_{V} )} & \notag\\
	& - {\sum_{m\in \mathcal{M}}{\theta_{m}( {{\sum_{k\in \mathcal{K}}{y_{km}f_{km} -}}F_{m}} )}}\,. \tag{54}
\end{flalign}
where $\boldsymbol{\lambda}$, $\boldsymbol{\mu}$ and $\boldsymbol{\theta}$ are the Lagrangian multipliers. The Lagrangian dual function (i.e., the objective dual problem) is then formulated as:
\begin{equation}
	\label{eq55}
	\mathcal{D}\left( {\boldsymbol{\lambda},\boldsymbol{\mu},\boldsymbol{\theta}} \right) = {\max{\mathcal{L}\left( {\mathbf{F},\boldsymbol{\lambda},\boldsymbol{\mu},\boldsymbol{\theta}} \right).}} \tag{55}
\end{equation}

Further, the object is to minimize the dual function over Lagrangian multiplier $\boldsymbol{\lambda}$, $\boldsymbol{\mu}$ and $\boldsymbol{\theta}$. Thus, the dual problem of $P1.2.2.1$ is given by:
\begin{flalign}
	\label{eq56}	
	~~~~&\min\limits \mathcal{D}(\boldsymbol{\lambda},\boldsymbol{\mu}, \boldsymbol{\theta}) & \notag\\
	&\qquad ~ \mbox{s.t.~}\boldsymbol{\lambda} \geq 0,~\boldsymbol{\mu} \geq 0,~\boldsymbol{\theta} \geq 0\,. & \tag{56}
\end{flalign}

According to Lemma 2, $P1.2.2.1$ is convex as ${{\partial^{2}U}/{\partial\mathbf{F}^{2}}} \leq 0$, Therefore, there exists a strictly feasible point, and Slater's condition holds, which leads to strong duality \cite{boyd2004convex}. Then, we can solve the primal problem $P1.2.2.1$ via the dual problem (55), which can be solved applying the gradient method. As the Lagrange function is differentiable, we can obtain the gradients of the Lagrange multipliers as:
\begin{flalign}
	\label{eq57}	
	\qquad&\frac{\partial\mathcal{L}\left( {\mathbf{F},\boldsymbol{\lambda},\boldsymbol{\mu}, \boldsymbol{\theta}} \right)}{{\partial\lambda}_{k}} = - {\sum\limits_{m\in \mathcal{M}}{y_{km}t_{km} + T_{k}^{max}}}\,; & \notag\\
	\qquad&\frac{\partial\mathcal{L}\left( {\mathbf{F},\boldsymbol{\lambda},\boldsymbol{\mu}, \boldsymbol{\theta}} \right)}{{\partial\mu}_{km}} = y_{km}f_{km} - F_{v}\,; & \notag\\
	\qquad& \frac{\partial\mathcal{L}\left( {\mathbf{F},\boldsymbol{\lambda},\boldsymbol{\mu}, \boldsymbol{\theta}} \right)}{\partial\theta_{m}} = - {\sum\limits_{k\in \mathcal{K}}{y_{km}f_{km} +}}F_{m}\,. &\tag{57}
\end{flalign}

Further, the Lagrangian multipliers are calculated iteratively as:
\setlength{\lineskip}{2.5pt}
\setlength{\lineskiplimit}{2.5pt}
\begin{flalign}
	\label{eq58}	
	\qquad&\lambda_{k}( {\psi + 1} )=\left[\lambda_{k}\left( j \right) + \varphi_{1}\frac{\partial\mathcal{L}}{\partial\lambda_{k}}\right]^{+}; & \notag\\
	\qquad&\mu_{km}( {\psi + 1} )=\left[\mu_{k}\left( j \right) + \varphi_{2}\frac{\partial\mathcal{L}}{\partial\mu_{km}}\right]^{+}; & \notag\\
	\qquad&\theta_{m}( {\psi + 1} )=\left[\theta_{m}\left( j \right) + \varphi_{3}\frac{\partial\mathcal{L}}{\partial\theta_{m}}\right]^{+}, &\tag{58}
\end{flalign}
where $\psi$ denotes the gradient number, $\varphi_{1}$, $\varphi_{2}$, $\varphi_{3}>0$ are the gradient steps. $[~]^{+}$ represents $\max (0,\cdot)$. The first-order derivative of $\mathcal{L}$ with respect to $f_{km}$ is:
\begin{flalign}
	\label{eq59}	
	 \frac{\partial\mathcal{L}\left( {\mathbf{F},\boldsymbol{\lambda},\boldsymbol{\mu}, \boldsymbol{\theta}} \right)}{\partial f_{km}} =& \frac{y_{km}\beta{\tilde{c}}_{k}f_{km}^{- 2}}{ln\left( 2 \right)T_{k}^{max}\left( 1 + \alpha a_{km} + \beta b_{km} \right)}  \notag\\
	 & + \frac{\lambda_{k}y_{km}{\tilde{c}}_{k}}{f_{km}^{2}} + {\mu_{km}y}_{km} - \theta_{m}y_{km}\,. \tag{59}
\end{flalign}

By setting (59) to zero, we can obtain $\mathbf{F}^{*} = \left\{ f_{km}^{*} \middle| \forall k \in \mathcal{K},\mathcal{~}m \in \mathcal{M} \right\}$.

\textit{(2) Offloading Ratio and Transmit Power: }We can obtain the optimal computation resource allocation under given $\boldsymbol{\varrho}$ and $\mathbf{P}$. On the contrary, the offloading ratio and transmit power problem for the given $\mathbf{P}$ is denoted as
\begin{flalign}
	\label{eq60}	
	~~~~P1.2.2.2:~&\max\limits_{\boldsymbol{\varrho},\mathbf{P}} U & \notag\\
	&\mbox{s.t.~(11),~(25b),~(25e)~and~(25f)}\,. & \tag{60}
\end{flalign}
\vspace{-5mm}
\begin{algorithm}[h]
	\caption{Joint Optimization of Energy Consumption and Task Processing Delay (JOET)} %算法的名字
	%\hspace*{0.02in} {\bf Input:} %算法的输入， \hspace*{0.02in}用来控制位置，同时利用 \\ 进行换行
%	\textbf{Input:} $\mathbf{x}$ and $\mathbf{y}$ \\
%	% \hspace*{0.02in} {\bf Output:} %算法的结果输出
%	\textbf{Input:} $\mathbf{x}^*$ and $\mathbf{y}^*$
	\begin{algorithmic}[1]
		\State \textbf{Initialization: } Let $\phi=1$; Initialize selection decisions $\mathbf{z}^{(0)}$, computation resources $\mathbf{F}^{(0)}$, offloading ratio $\boldsymbol{\varrho}^{(0)}$ and  transmit power $\mathbf{P}^{(0)}$. % \State 后写一般语句
		\Repeat
		\State /* Optimization of Selection Decisions */
		\State Under given $\mathbf{F}^{(0)}$, $\boldsymbol{\varrho}^{(\phi-1)}$ and $\mathbf{P}^{(\phi-1)}$, derive $\mathbf{z}^{(\phi-1)}$\par \noindent \quad $~$ by Solving $P1.1$ with Algorithm 1.
		\State /* Joint Optimization of Resource Allocation, \par Offloading Ratio and Transmit Power */
		\If{${\sum_{m \in \mathcal{M}}y_{km}} = 1$}
		\State /* Offloading to the server by the vehicle relay */
		\State Let $\psi=1$; Initialize Lagrangian multipliers \par ~~$\,\boldsymbol{\lambda}(0),\boldsymbol{\mu}(0)$ and $\boldsymbol{\theta}(0)$;
		\Repeat
		\State Update Lagrangian multipliers $\boldsymbol{\lambda}(\psi),\boldsymbol{\mu}(\psi)$ \par \hskip 2.5em  and $\boldsymbol{\theta}(\psi)$ based on (58);
		\State Then determine $\mathbf{F}^{(\phi)}$ via calculating \par \hskip 2.5em $\partial\mathcal{L}\left( \mathbf{F},\boldsymbol{\lambda},\boldsymbol{\mu}, \boldsymbol{\theta} \right) / ( \partial f_{km})=0$ based on (59)
		\State $\psi = \psi +1$
		\Until{Convergence}
		\EndIf
		\State Derive Offloading Ratio $\boldsymbol{\varrho}^{(\phi )}$ and Transmit \par \noindent \quad $\;$ Power $\mathbf{P}^{(\phi )}$  by (48) and (50) respectively.
		\State $\phi =\phi +1$
		\Until{Convergence}
	\end{algorithmic}
\end{algorithm}

According to Lemma 1, $P1.2.2.2$ is non-convex problem. The process of solving this problem is the same as that of $P1.2.1$ in Section \Rmnum{4}-B1. Due to limited space of this paper, we will not repeat it here. Referring to Section \Rmnum{4}-B1, we can get the optimal solution  $\boldsymbol{\varrho}^{*}$ and $\mathbf{P}^{*}$ in same way.

\subsection{JOET: Joint Optimization for Energy Consumption and Task Processing Delay}
According to the discussion of previous subsections, we can efficiently solve subproblems of $P1$ in a distributed manner. Specifically, in order to minimize energy consumption and task processing delay, we decouple $P1$ into optimization of selection decisions and optimization of computation resources, offloading ratio and transmit power. That is, under given $\mathbf{P}$, $\boldsymbol{\varrho}$ and $\mathbf{P}$, the VNO derives $\mathbf{z}$ for each $u_k$. Further, $\mathbf{F}$ is obtained under given $\mathbf{z}$, $\boldsymbol{\varrho}$ and $\mathbf{P}$ by Lagrangian method and then $\boldsymbol{\varrho}$ and $\mathbf{P}$, is derived under given $\mathbf{z}$ and $\mathbf{F}$, and repeat this process until convergence.

In Algorithm 2, since the computational complexity for solving problem $P1.1$ is only polynomial in convergence factor $\epsilon$, the computational complexity required to solve P1.1 is $\mathcal{O}(\log\left( 1/\epsilon \right))$. For $j$ iterations, the complexity of the inner loop of Algorithm 2 is $\mathcal{O}(j)$. Finally, the total complexity of the Algorithm 2 is $\mathcal{O}(\phi( 3 + \psi + \log(1/\epsilon)))$. Different from Ref. \cite{dai2018joint}, offloading the load is only the second subproblem of this paper. In addition, this paper not only considers the return result in detail, but also optimizes the energy consumption.

\section{PERFORMANCE ANALYSIS}
This section demonstrates extensive simulations to evaluate the performance of the proposed JOET scheme.
\subsection{Simulation Environment}
Considering a 3km$\times$3km square area, which consists of 25 small areas. There are 6 RSUs evenly located in the square area. Each RSU is equipped with a VEC server. The total computation resources of the VEC servers are $F_m=5$GHz, $\forall m\in \mathcal{M} $. Since in-vehicle applications also require computation resources, it is assumed that free resources $F_V$=1.0GHz and $\bar{v}=60$km/h for each vehicle; In the 15 small areas on the left, each small area has two IoTDs, and in the 10 small areas on the right, each small area has a IoTD, so K=40. Each IoTD has a computation task in each time slot $n\in \mathcal{N}$, where $N=60,\;\delta_{t}=1$s. In all computation tasks, input data size, output data size, local computation resources and the required computation resources are evenly distributed in the range of U[10,640]KB, U[5,300]KB, U[0.1,1]GHz and U[0.2,2]GHz, respectively. Besides, the maximum processing delay and the maximum energy consumption are 1.15 times of the case that the task is computed locally, and the convergence factor $\epsilon=0.1$. The other parameters are summarized in TABLE \Rmnum{1}.

To better evaluate the effectiveness of JOET scheme, we introduce several other benchmark schemes as follows.

\textit{1) No Vehicle design (NoVeh):} In this scheme, computation tasks of IoTDs can only be computed by themselves or offloaded to the VEC server without vehicle assistance;

\textit{2) No VEC server design (NoVEC):} In this scheme, computation tasks of IoTDs can only be computed by themselves or offloaded to the vehicle.

\textit{3) Only relaying design (OnlyR):} In this scheme, the vehicle can only act as a relay to assist the task transmission from the IoTD to the VEC server \cite{huang2020result}.

\textit{4) Selection Optimization (SO):} In this scheme, the best VEC server with maximal system utility is selected by each vehicle under a given computation resources ($f_{km}=1.2$G) and max transmit power. Then the server allocates evenly the remaining resources to $u_k$s selecting the server \cite{dai2018joint}.

\textit{5) Computation Resources and Transmit Power (CRTP):} In this scheme, computation resources and transmit power are jointly optimized. Herein, $u_k$ offload the task to the nearest VEC servers as CRTP does not consider VEC server selection. This scheme is similar to Ref. \cite{yu2016joint}, which jointly optimizes resource allocation and offloading ratio.

\setlength{\lineskip}{2.5pt}
\setlength{\lineskiplimit}{2.5pt}
\textit{6) Exhaustive Search  Method (ESM):} In this scheme, at least $( M + 1)^{K}( \frac{F_{m}}{\mathrm{\Delta}_{f}})^{K}( \frac{F_{m}}{\mathrm{\Delta}_{p}})^{K}( \frac{1}{\mathrm{\Delta}_{\varrho}})^{K} $ combinations of offloading schemes are evaluated to find out the optimal solution with highest utility. $\mathrm{\Delta}_{f},\mathrm{\Delta}_{p}$ and $\mathrm{\Delta}_{\varrho}$ are the stride step in exhaustion. Because of the enormous computational complexity, it takes a huge amount of time to run this method, so we only conduct a experiment via pruning in Section \Rmnum{5}-G. \vspace{-1.5mm}
\subsection{Evaluation measure}
\setlength{\lineskip}{2.5pt}
\setlength{\lineskiplimit}{2.5pt}
Ref. \cite{dai2018joint} adopts the number of tasks to evaluate the load balancing of servers. However, when the differences (required computation resources, maximum task processing delay, etc.) among each task vary greatly, this method is difficult to measure load balancing between servers. Therefore, we propose an approach which adopts the variance of the average resource utilization efficiency to better evaluate load balancing, where the average resource utilization efficiency is denoted as $\rho_{m} = \frac{1}{M_{m}}{\sum_{i = 1}^{M_{m}}\frac{{\tilde{c}}_{k}}{f_{km}}}$. Further, the variance is calculated by:\\
\begin{equation}
	\label{eq61}
	\mathrm{s}^{2} = \frac{1}{M}{\sum_{m\in \mathcal{M}}{\left( \rho_{m} - {\bar{\rho}}_{m} \right)^{2}.}} \tag{61}
\end{equation}

In particular, if $\mathrm{s}^2$ is smaller, it indicates that the total resources required for tasks offloaded to each server are closer to the average, which makes the most fair load allocation.
\begin{figure*}[tbp!]
	\centering
	\setlength{\abovecaptionskip}{-3pt}
	\setlength{\belowcaptionskip}{-8pt}
	\begin{center}
		\captionsetup{name={},justification=centering}
		\subfloat[$~\alpha=1,~\beta=1$]{
			\begin{minipage}[t]{0.33\linewidth}
				%				\centering
				\includegraphics[width =0.92\linewidth]{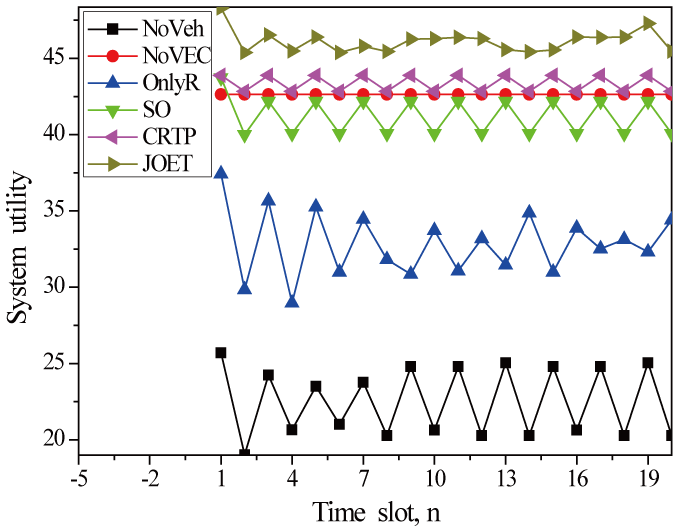}
				\label{fig:side:3a}
			\end{minipage}
		}
		\subfloat[$~\alpha=100,~\beta=1$]{
			\begin{minipage}[t]{0.33\linewidth}
				%				\centering
				\includegraphics[width =0.92\linewidth]{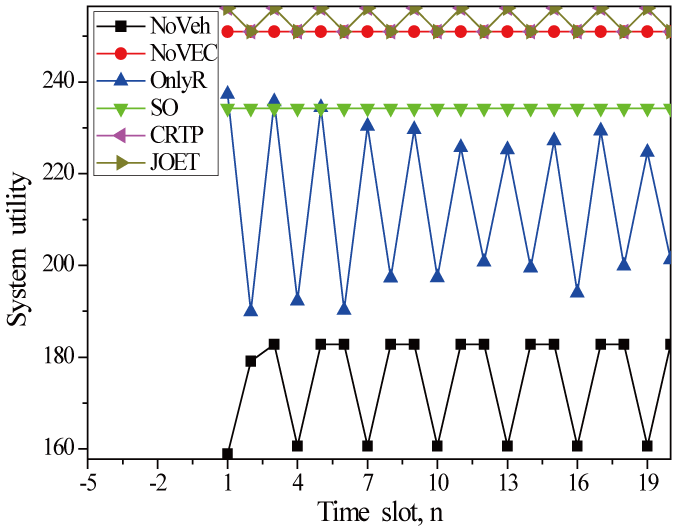}
				\label{fig:side:3b}
			\end{minipage}
		}
		\subfloat[$~\alpha=1,~\beta=100$]{
			\begin{minipage}[t]{0.33\linewidth}
				%				\centering
				\includegraphics[width =0.92\linewidth]{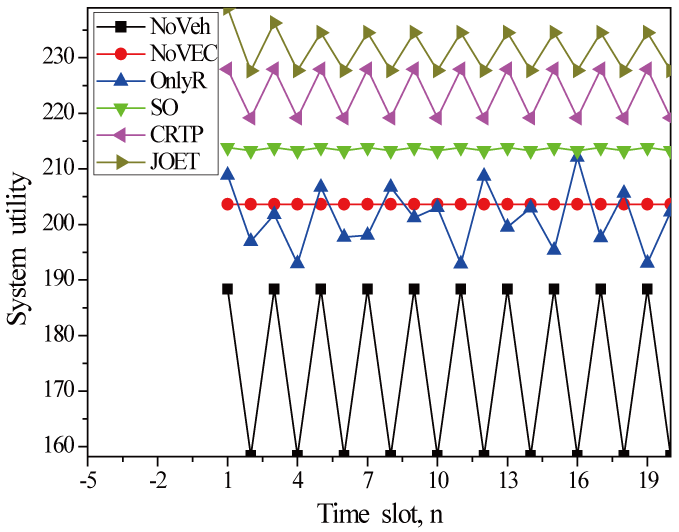}
				\label{fig:side:3c}
			\end{minipage}
		}
	\end{center}
	\caption{Comparison of system utility with respect to time slot $n$ under six schemes with  $K=40$ and $M=6$ (OnlyR is from Ref. \cite{huang2020result}; SO is from Ref.  \cite{dai2018joint}; CRTP is from Ref. \cite{yu2016joint}).}
	%	\captionsetup 
\end{figure*}
\begin{figure*}[tbp!]
	\centering
	\setlength{\abovecaptionskip}{-3pt}
	\setlength{\belowcaptionskip}{-8pt}
	\begin{center}
		\captionsetup{name={},justification=centering}
		\subfloat[$~\alpha=1,~\beta=1$]{
			\begin{minipage}[t]{0.33\linewidth}
				%				\centering
				\includegraphics[width =0.92\linewidth]{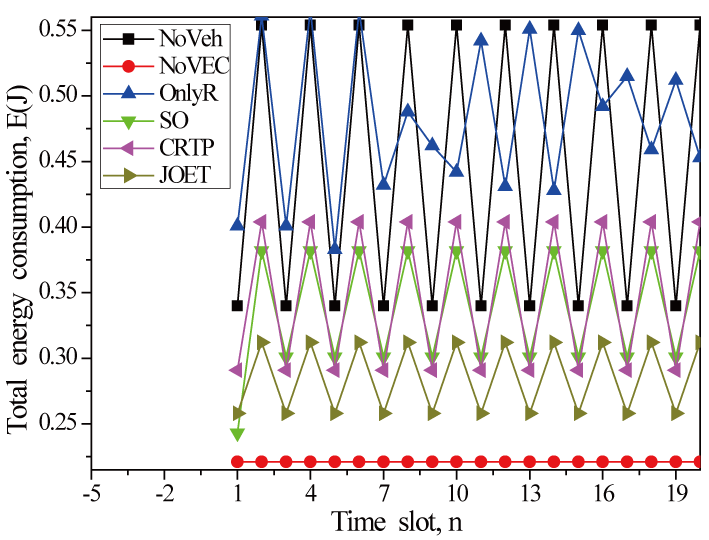}
				\label{fig:side:4a}
			\end{minipage}
		}
		\subfloat[$~\alpha=100,~\beta=1$]{
			\begin{minipage}[t]{0.33\linewidth}
				%				\centering
				\includegraphics[width =0.92\linewidth]{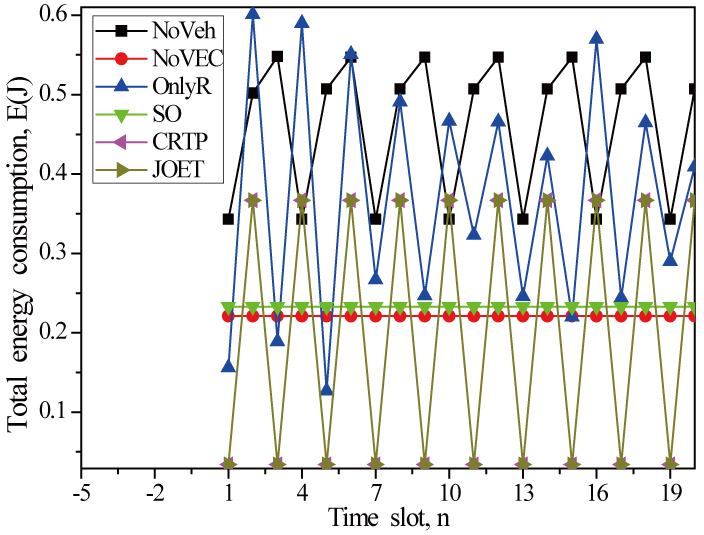}
				\label{fig:side:4b}
			\end{minipage}
		}
		\subfloat[$~\alpha=1,~\beta=100$]{
			\begin{minipage}[t]{0.33\linewidth}
				%				\centering
				\includegraphics[width =0.92\linewidth]{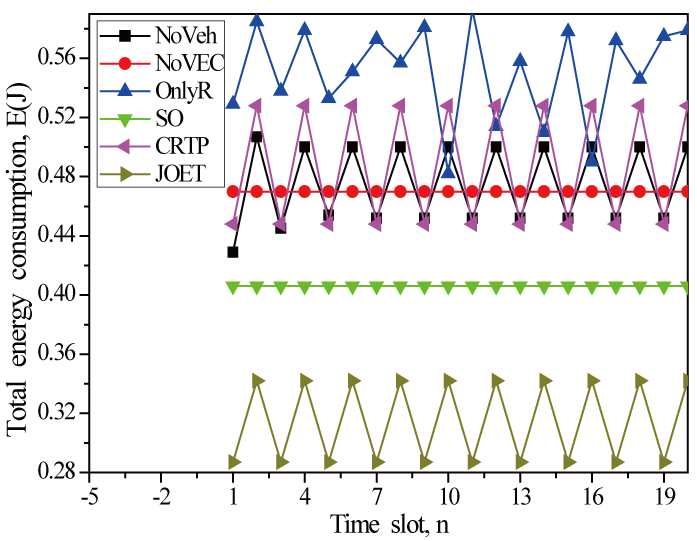}
				\label{fig:side:4c}
			\end{minipage}
		}
	\end{center}
	\caption{Comparison of the average energy consumption under six schemes with $K=40$ and $F_m=5$ GHz (OnlyR is from Ref. \cite{huang2020result}; SO is from Ref. \cite{dai2018joint}; CRTP is from Ref. \cite{yu2016joint}).}
	%	\captionsetup 
\end{figure*}
\begin{figure*}[tbp!]
	\centering
	\setlength{\abovecaptionskip}{-3pt}
	\setlength{\belowcaptionskip}{-8pt}
	\begin{center}
		\captionsetup{name={},justification=centering}
		\subfloat[$~\alpha=1,~\beta=1$]{
			\begin{minipage}[t]{0.33\linewidth}
				%				\centering
				\includegraphics[width =0.92\linewidth]{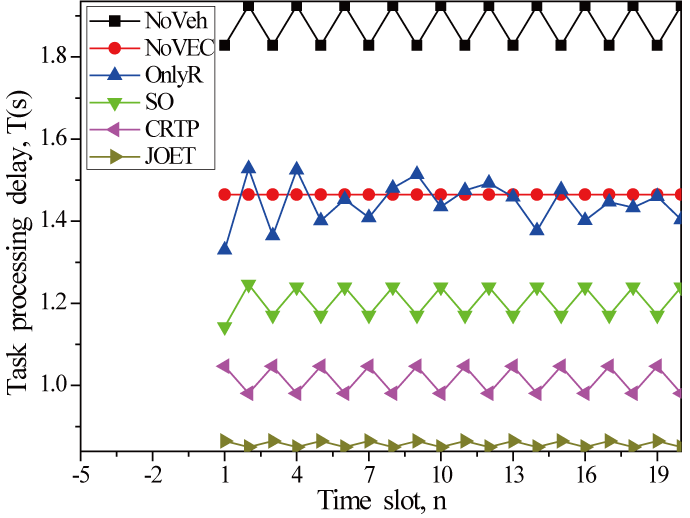}
				\label{fig:side:5a}
			\end{minipage}
		}
		\subfloat[$~\alpha=100,~\beta=1$]{
			\begin{minipage}[t]{0.33\linewidth}
				%				\centering
				\includegraphics[width =0.92\linewidth]{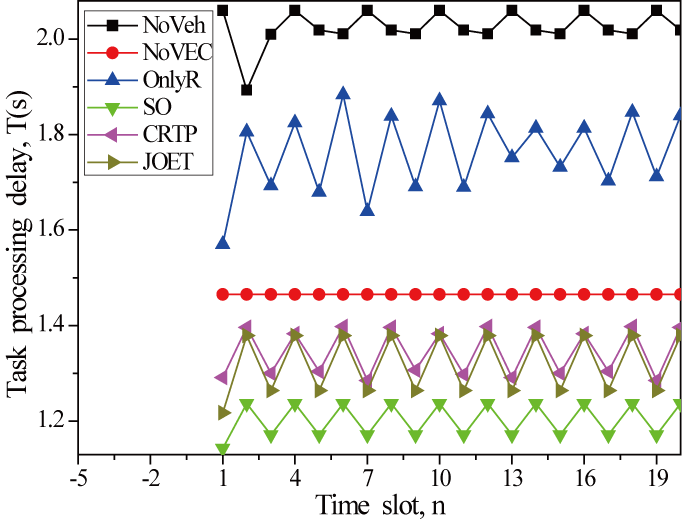}
				\label{fig:side:5b}
			\end{minipage}
		}
		\subfloat[$~\alpha=1,~\beta=100$]{
			\begin{minipage}[t]{0.33\linewidth}
				%				\centering
				\includegraphics[width =0.92\linewidth]{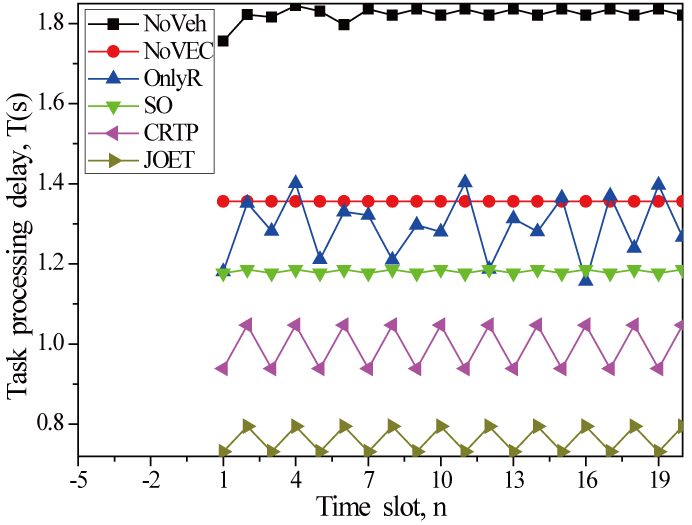}
				\label{fig:side:5c}
			\end{minipage}
		}
	\end{center}
	\caption{Comparison of the average task processing delay under six schemes with $K=40$ and $F_m=5$ GHz (OnlyR is from Ref. \cite{huang2020result}; SO is from Ref.  \cite{dai2018joint}; CRTP is from Ref. \cite{yu2016joint}).}
	%	\captionsetup 
\end{figure*}
\subsection{System Utility}
Firstly, we compare system utility with respect to time slot under six schemes. Fig. 3 shows JOET always achieves the greatest system utility. In Fig. 3(a), the average system utility of JOET and the other six schemes are 46.068, 43.357, 41.203, 32.838, 42.636 and 22.489 respectively. System utility of JOET is 6.25$\%$ higher than CRTP. In Fig. 3(b), six schemes are 174.78, 250.96, 213.12, 234.26, 253.52 and 253.53, respectively. System utility of JOET is basically same as that of CRTP. This is because JOET and CRTP in order to minimize energy consumption, and the both offload tasks to vehicles or servers as much as possible and optimize transmit power. In Fig. 3(c), six schemes are 173.375, 203.64, 201.20, 213.57 223.55 and 231.41, respectively. It can observed that the result is similar to Fig. 3(b) and other five schemes underperformer JOET. This is because these schemes often ignore one optimization object when they focus on optimizing another optimization object (energy consumption or task processing delay). In addition, system utility of six schemes fluctuates as time goes on. This is because all schemes only consider the maximum system benefit of the current time slot, but do not consider random arrival of tasks and vehicles and dynamic energy harvesting process. We will consider these important factors in future research works.

\subsection{Energy Consumption and Task Processing Delay}
In this section, we compare energy consumption and task processing delay of six schemes. Fig. 4(a) shows that the average energy consumption of  NoVeh, NoVEC, OnlyR, SO, CRTP and JOET are 0.447, 0.221, 0.472, 0.339, 0.348 and 0.285, respectively. The proposed JOET reduces $15.93\%$ compared with SO. Note that, although NoVEC achieves the lowest energy consumption, the task processing delay is greatly increased (as shown in Fig. 4(a)). This is because NoVEC cannot allocate more computation resources without VEC to effectively reduce the task processing delay. Therefore, NoVEC cannot satisfies the latency-sensitive tasks. Similar to the system utility experiments, the system utility of JOET and CRTP are similar because both optimize jointly offloading ratio and transmit power. In addition, OnlyR and NoVeh consume more energy than SO because these schemes cannot sustainably offload tasks to the vehicle for computing, which increases task processing delay. When the time taken to offloading to the server exceeds the maximum task processing delay $T^{max}$, the task will be computed locally. In this case, energy consumption caused by local computation is close to the maximum energy consumption. Therefore, although OnlyR and NoVeh optimize offloading ratio and task processing delay, they are underperform SO in terms of energy consumption.

In Fig. 5(a), the average task processing delay of NoVeh, NoVEC, OnlyR, SO, CRTP and JOET are 1.876, 1.465, 1.476, 1.443, 1.204 and 1.014 respectively, where JOET reduces task processing delay by $15.78\%$ compared to CRTP. In particular, task processing delay of NoVeh is always the greatest. This is because the task cannot be amplified and transmitted to the server through the vehicle relay without vehicle assistance, which leads to increase transmission delay. In summary, JOET minimizes task processing delay while also minimizing energy consumption through joint optimization of selection decisions, resource allocation, offloading ratio and transmit power adjustment (especially in Fig. 4(c) and Fig. 5(c)).
\begin{figure*}[tbp!]
	\centering
	\setlength{\abovecaptionskip}{-3pt}
	\setlength{\belowcaptionskip}{-8pt}
	\begin{center}
		\captionsetup{name={},justification=centering}
		\subfloat[$~\alpha=1,~\beta=1$]{
			\begin{minipage}[t]{0.33\linewidth}
				\includegraphics[width =0.92\linewidth]{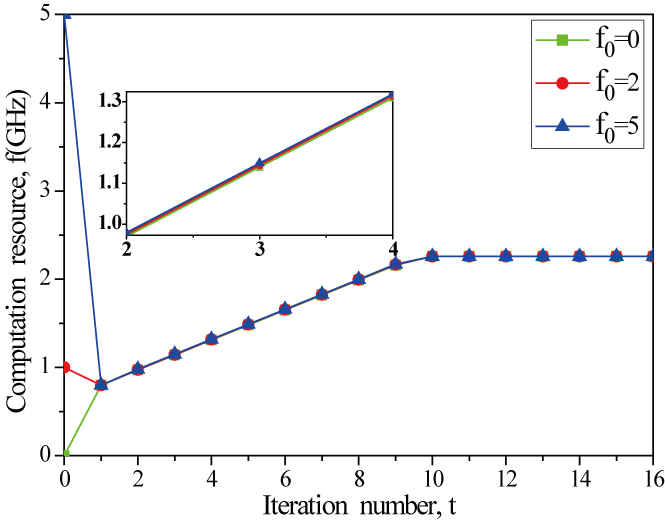}
				\label{fig:side:6a}
			\end{minipage}
		}
		\subfloat[$~\alpha=100,~\beta=1$]{
			\begin{minipage}[t]{0.33\linewidth}
				%				\centering
				\includegraphics[width =0.92\linewidth]{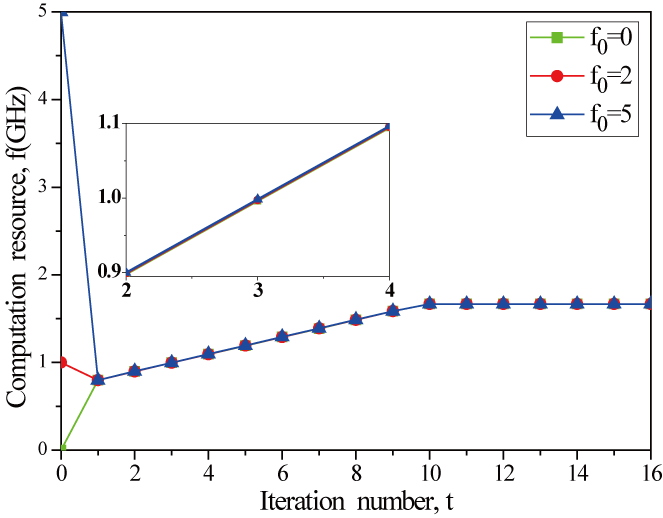}
				\label{fig:side:6b}
			\end{minipage}
		}
		\subfloat[$~\alpha=1,~\beta=100$]{
			\begin{minipage}[t]{0.33\linewidth}
				%				\centering
				\includegraphics[width =0.92\linewidth]{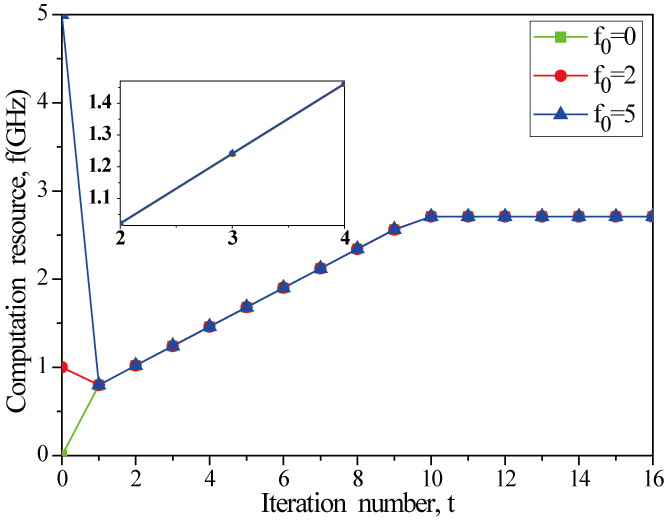}
				\label{fig:side:6c}
			\end{minipage}
		}
	\end{center}
	\caption{The variation of computation resources allocated to the task requiring 1.8 GHz resources under different initial point with $K=40$ and $F_m= 5$ GHz.}
	%	\captionsetup 
\end{figure*}
\begin{figure*}[tbp!]
	\centering
	\setlength{\abovecaptionskip}{-3pt}
	\setlength{\belowcaptionskip}{-8pt}
	\begin{center}
		\captionsetup{name={},justification=centering}
		\subfloat[$~\alpha=1,~\beta=1$]{
			\begin{minipage}[t]{0.33\linewidth}
				%				\centering
				\includegraphics[width =0.92\linewidth]{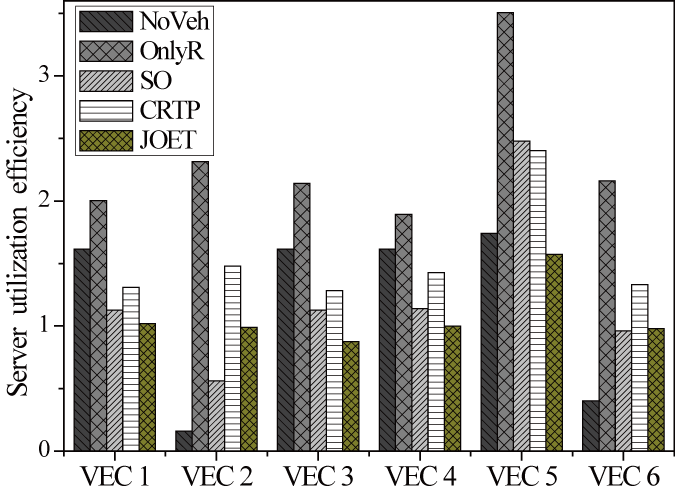}
				\label{fig:side:7a}
			\end{minipage}
		}
		\subfloat[$~\alpha=100,~\beta=1$]{
			\begin{minipage}[t]{0.33\linewidth}
				%				\centering
				\includegraphics[width =0.92\linewidth]{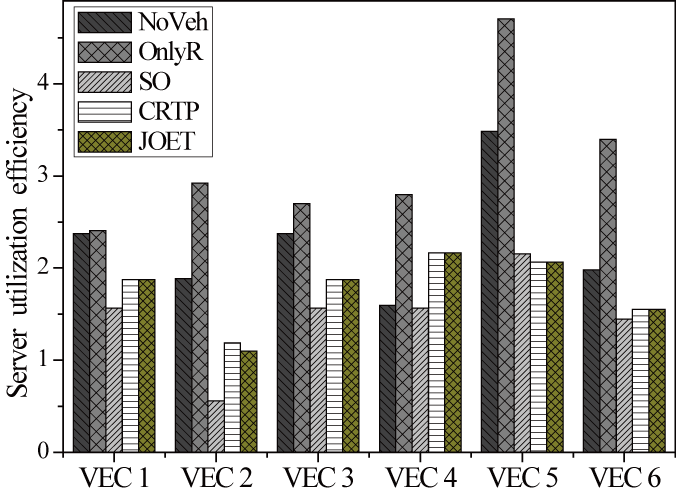}
				\label{fig:side:7b}
			\end{minipage}
		}
		\subfloat[$~\alpha=1,~\beta=100$]{
			\begin{minipage}[t]{0.33\linewidth}
				%				\centering
				\includegraphics[width =0.92\linewidth]{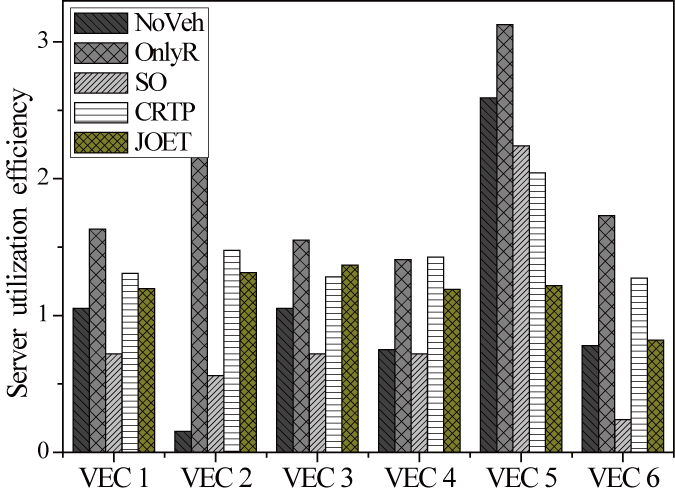}
				\label{fig:side:7c}
			\end{minipage}
		}
	\end{center}
	\caption{Comparison of load balancing under five schemes with $K=40$ and $F_m= 5$ GHz (OnlyR is from Ref. \cite{huang2020result}; SO is from Ref. \cite{dai2018joint}; CRTP is from Ref. \cite{yu2016joint}).}
	%	\captionsetup 
\end{figure*}
\subsection{Convergence}
In the above two sections, we have proved that the proposed JOET outperformers the other five schemes. Furthermore, in this section, we take a convergence experiment at different initial points to verify the feasibility of JOET. In the experiment, we conduct a task requiring 1.8 GHz resources as an example. As shown in Fig. 6, it illustrates convergent evolution of the inner loop of Algorithm 2, i.e., the loop from step 9 to step 13. In Fig. 6(a), when $\alpha=1$ and $\beta=1$, computation resources allocated to the task after convergence reach 2.259GHz. Since the required resources for each task are greater than local resources, tasks that require less computation resources can be offloaded to the vehicle to greatly reduce task processing delay. Thus, the system allocates server resources to tasks that require more computation resources. In Fig. 6(b), it can be concluded that the allocated resources are only 1.668 after convergence. In order to minimize energy consumption, it is necessary to offload tasks to vehicles or servers as sustainable as possible because the communication-related energy consumption is much smaller than the computation-related energy consumption. When more tasks are assigned to the server to obtain more resources, the server can only reduce resources allocated to each task. This is different from Fig. 6(c) in that the gap between local computation delay and offloading processing delay is smaller than the gap between the computation-related energy consumption and the communication-related energy consumption. It can also be concluded that resources allocated to tasks can basically reach converge within ten iterations and converges almost simultaneously under different initial points. Thus, convergence of JOET algorithm is similar to the result of Fig. 6. Finally, it can be concluded that JOET can effectively solve the problem $P1$ based on the above results.

\subsection{Load Balancing}
In order to effectively balance the server load, JOET integrates load balancing into the task offloading problem. Different from Ref. \cite{dai2018joint}, which visually presents the optimal scheme achieving the fairest load, we use the variance of the average resource utilization efficiency reflects the server load truly. As shown in Fig. 7(a), $\mathrm{s}^2$ of NoVeh, OnlyR, So, CRTP and JOET are 0.422, 0.292, 0.352, 0.154 and 0.052 respectively. JOET performs the fairest load so that the variance is the smallest and reduce 10.2$\%$ variance than CRTP. In addition, the resource utilization of OnlyR is highest. This is because the vehicle is only used as a relay to assist tasks so that more tasks are offloaded to the server for computing, which results in the highest resource utilization in OnlyR. Similar to system utility experiments, the server resource utilization of JOET and CRTP are close in Fig. 7(b). This is because the both focus on minimizing energy consumption by jointly optimizing offloading ratio and transmit power. Note that, although NoVeh and OnlyR also optimize offloading ratio and transmit power. However, NoVeh cannot access more serves without vehicle relay; OnlyR cannot offloading the task to the vehicle for computing. Therefore, the both make a unbalanced assignment of load. According to the above overview, JOET makes the fairest task assignment, which makes the load of each server as even as possible. This is because JOET does its best by integrating load balancing into VEC selection decisions. On the contrary, SO does not consider balancing server load. Although NoVeh, OnlyR and CRTP consider balancing the server load, NoVeh cannot get vehicle assistance; OnlyR only considers the vehicle as a relay; CRTP is sensitive to outliers because it only selects nearby server. Through the above experiments, the proposed JOET effectively offloads tasks with consideration of load balancing.
\begin{figure}[!t]
	\centering
	\setlength{\abovecaptionskip}{5pt}
	\includegraphics[width=0.96\linewidth]{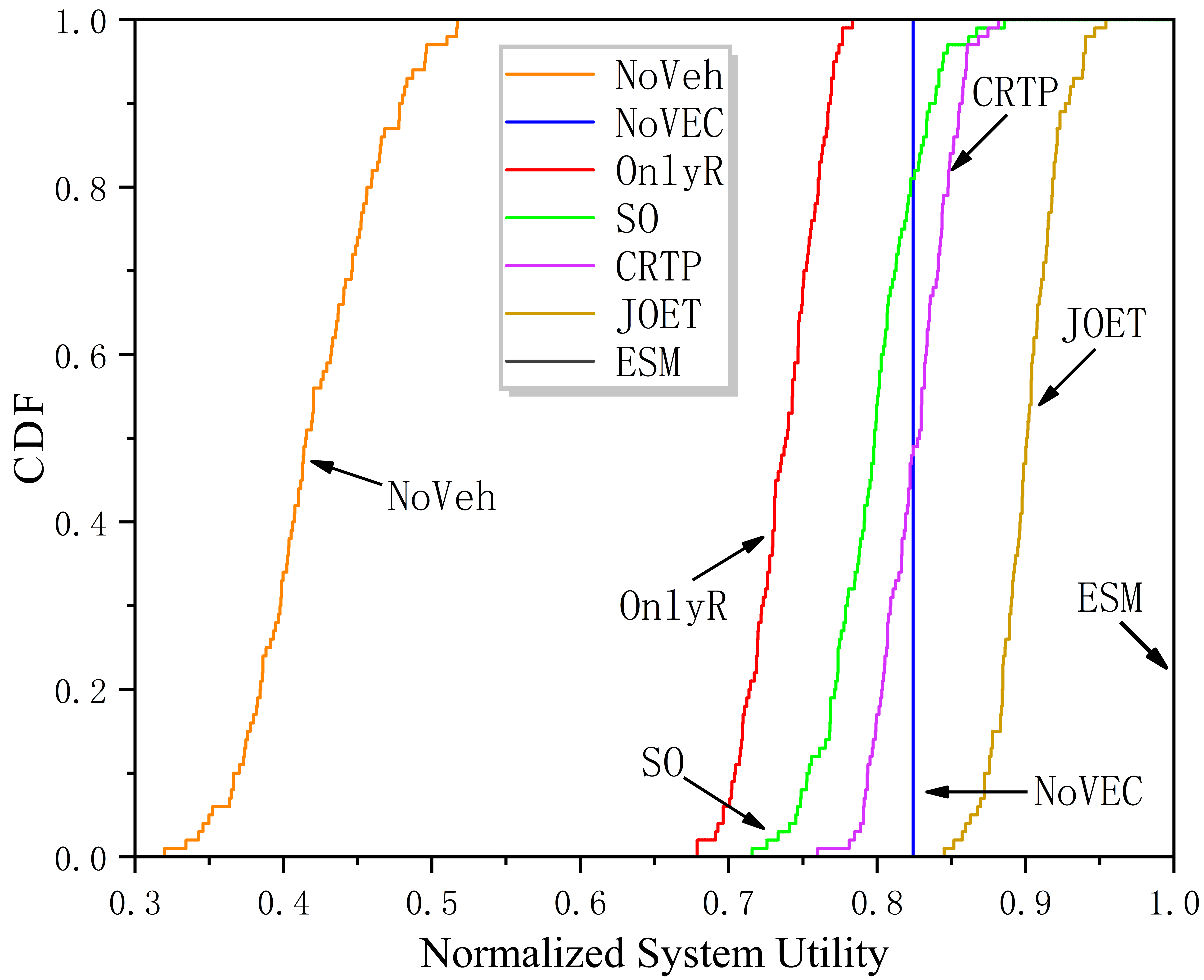}
	\caption{\label{fig8} Variation of system utility under different schemes with $K=40$ and $M=6$. (OnlyR is from Ref. \cite{huang2020result}; SO is from Ref. \cite{dai2018joint}; CRTP is from Ref. \cite{yu2016joint}).}
\end{figure}
\subsection{Engineering Applications}
We will focus on dynamics and randomness making vehicle-assisted VEC more complex in future work. But before that, we study a preliminary random task generation. In the previous experiments, the tasks generated by IoTDs are fixed. In order to evaluate the generalization ability of these offloading schemes, we analyze the performance of each schemes across different simulation runs that randomly generate different tasks for each IoTD. In simulation run, the system utility of each scheme is normalized, which is presented as a ratio of the system utility to the optimal one obtained by ESM. Obviously, the ratio cannot be greater than 1, and the ESM performs 1 in each simulation run. We adopt the Cumulative Distribution Function (CDF) to present the distribution of the ratios across all the simulation runs \cite{zhan2020mobility} when all tasks from IoTDs are randomly generated. As shown in Fig. 8, the utility ratio of NoVEC is always equal to 0.824. This is because NoVEC always offloads tasks to vehicles and thus achieves the same system utility, which is consistent with Fig .3. Similarly, NoVeh achieves the lower system utility because of random task assignment. In Fig. 3, SO and CRTP are close to JOET. However, due to the random task assignment, SO without optimization of resources cannot make better selection decisions, and CRTP that offloads nearby tasks cannot meet more requirement of tasks. JOET performs the better generalization ability by jointly optimizing selection decisions, resource allocation, offloading ratio and transmit power adjustment.

\section{Conclusion and Future Work}
In this paper, we investigated a sustainable vehicle-assisted VEC system, where the vehicle not only can compute the latency-sensitive tasks, but also act as a relay to help tasks from IoTDs to VEC servers. We proposed integrating load balancing with the offloading problem to efficiently complete tasks and further formulate the problem as a mix-integer non-linear optimization problem. By employing decomposition, the problem is decoupled into two separable subproblems, and finally solve the original problem by solving subproblems iteratively. In summary, we proposed a joint optimization energy consumption and task processing delay scheme, called JOET, which aims to sustainably reduce the total energy consumption and task processing delay via jointly optimizing selection decisions, computation resource allocation, offloading ratio and transmit power adjustment. Finally, we conducted extensive simulation experiments. The results show that our proposed JOET minimizes energy consumption and task processing delay while performing a fairer task assignment than other schemes. 

For the future work, because the vehicle-assisted system does not consider randomization and dynamics, that is, random arrival of tasks and vehicles and dynamic energy harvesting process, which makes the optimization problem more arduous, we will pour attention to dynamics and randomization to make the offloading scheme in sustainable vehicle-assisted VEC system more feasible.

% if have a single appendix:
%\appendix[Proof of the Zonklar Equations]
% or
%\appendix  % for no appendix heading
% do not use \section anymore after \appendix, only \section*
% is possibly needed

% use appendices with more than one appendix
% then use \section to start each appendix
% you must declare a \section before using any
% \subsection or using \label (\appendices by itself
% starts a section numbered zero.)
%

\appendices
\section{Proof OF THE LEMMA 1}
The second-order derivative of $U$ with respect to $\boldsymbol{\varrho}$ is:
\begin{flalign}
	\label{eq62}
	\frac{\partial^{2}U}{\partial\boldsymbol{\varrho}^{2}} = & - \frac{1}{\ln\left( 2 \right)}\frac{\frac{\beta}{T_{k}^{max}}~\omega_{3}'}{\omega_{0}} - \frac{1}{\ln( 2 )}( \frac{\frac{I_{k}}{R_{k}^{k\rightarrow v}}\omega_{1} + \omega_{2} + \omega_{3}}{\omega_{0}})^{2}, \tag{62}
\end{flalign}
where $\omega_1,\;\omega_2$ and $\omega_3$ are defined in (46), $\omega_{3}' = \frac{( 2 + \omega_{3} ){\tilde{c}}_{k}\bar{v}}{( {1 + \frac{p_{v}\xi_{1}}{d_{vv}^{2}}} )R_{k}^{v\rightarrow v}{d_{vv}^{3}F}_{V}} \times ( \frac{4B_{0}p_{v}\xi_{1}}{\ln(2)} - ( 3d_{vv}^{2} + p_{v}\xi_{1} )R_{k}^{v\rightarrow v} )$.

Obviously, $d_{vv} > 1$, but $\frac{4B_{0}p_{v}\xi_{1}}{\ln\left( 2 \right)} - \left( {3d_{vv}^{2} + p_{v}\xi_{1}} \right)R_{k}^{v\rightarrow v}$ is indefinite. Further,
\begin{flalign}
	\label{eq63}
	\frac{\partial^{2}U}{\partial\boldsymbol{\varrho}\partial\mathbf{P}} = \frac{\partial^{2}U}{\partial\mathbf{P}\partial\boldsymbol{\varrho}} = & - \frac{\omega_{0}^{- 2}}{\ln(2)}( \frac{\alpha I_{k}}{\eta E_{k}^{max}R_{k}^{k\rightarrow v}} - \omega_{1}\omega_{4} ) &\notag \\
	& \times [ \omega_{0} + \varrho_{k}( \frac{I_{k}\omega_{1}}{R_{k}^{k\rightarrow v}} + \omega_{2} ) ]\,, & \tag{63}
\end{flalign}
where $\omega_{4} = \frac{I_{k}\frac{\xi_{0}}{d_{kv}^{2}}B_{0}( R_{k}^{k\rightarrow v} )^{- 2} }{{\ln( 2 )}( {1 + \frac{p_{k}\xi_{0}}{d_{kv}^{2}}})}$. Further, 
\begin{flalign}
	\label{eq64}
	\setlength{\belowdisplayskip}{6pt}
	\frac{\partial^{2}U}{\partial\mathbf{P}^{2}} = & - \frac{1}{\ln\left( 2 \right)}\frac{\omega_{4}\varrho_{k}( \omega_{1}\omega_{5} - \frac{2\alpha}{\eta E_{k}^{max}} )}{\omega_{0}} &\notag \\
	& - \frac{1}{\ln( 2 )}( \frac{\frac{\alpha}{\eta E_{k}^{max}}\frac{{\varrho_{k}I}_{k}}{R_{k}^{k\rightarrow v}} - \omega_{1}\omega_{4}\varrho_{k}}{\omega_{0}})^{2} , & \tag{64}
\end{flalign}
where $\omega_{5} = \frac{\xi_{0}}{d_{kv}^{2}( {1 + \frac{p_{k}\xi_{0}}{d_{kv}^{2}}} )}( {1 + \frac{2B_{0}}{{\ln( 2 )}R_{k}^{k\rightarrow v}}} )$. Then,
\begin{flalign*}
	\label{eq65-1}
	&\frac{\partial^{2}U}{\partial\boldsymbol{\varrho}^{2}}\frac{\partial^{2}U}{\partial\mathbf{P}^{2}} - \frac{\partial^{2}U}{\partial\boldsymbol{\varrho}\partial\mathbf{P}}\frac{\partial^{2}U}{\partial\mathbf{P}\partial\boldsymbol{\varrho}} = \frac{1}{( {\ln(2)} \omega_{0}^{2} )^{2} }   \notag \\
	&\times \left[ \frac{\beta\omega_{0}\omega_{3}'\varrho_{k}}{T_{k}^{max}} \times ( \omega_{0}\omega_{4}( \omega_{1}\omega_{5} - \frac{2\alpha}{\eta E_{k}^{max}} ) + \frac{\alpha}{\eta E_{k}^{max}}\frac{\varrho_{k}I_{k}}{R_{k}^{k\rightarrow v}} \notag \right. \\ 
	& -\omega_{1}\omega_{4}\varrho_{k}) + ( \frac{I_{k}\omega_{1}}{R_{k}^{k\rightarrow v}} + \omega_{2} + \omega_{3})^{2}( \omega_{0}\omega_{4}( \omega_{1}\omega_{5} - \frac{2\alpha}{\eta E_{k}^{max}} ) \notag \\
	& + \frac{\alpha}{\eta E_{k}^{max}}\frac{{\varrho_{k}I}_{k}}{R_{k}^{k\rightarrow v}} - \omega_{1}\omega_{4}\varrho_{k} ) - ( \frac{I_{k}}{R_{k}^{k\rightarrow v}}\omega_{1} + \omega_{2})^{2} \notag \\
	& \times ( \frac{\alpha}{\eta E_{k}^{max}}\frac{\varrho_{k}I_{k}}{R_{k}^{k\rightarrow v}} - \omega_{1}\omega_{4}\varrho_{k})^{2} - 2\omega_{0}\varrho_{k}( \frac{I_{k}}{R_{k}^{k\rightarrow v}}\omega_{1} + \omega_{2} )  \notag
\end{flalign*}
%\begin{flalign}
%	\label{eq65-2}
%
%\end{flalign}
\begin{equation}
	\label{eq65-3}
	\setlength{\belowdisplayskip}{5pt}
	\left. \times ( \frac{\alpha I_{k}}{\eta E_{k}^{max}R_{k}^{k\rightarrow v}} - \omega_{1}\omega_{4})^{2} - \omega_{0}^{2}( \frac{\alpha I_{k}}{\eta E_{k}^{max}R_{k}^{k\rightarrow v}} - \omega_{1}\omega_{4})^{2} \right]\,. \tag{65}
\end{equation}

According to the Section \Rmnum{5}-A and Table \Rmnum{1}, we have $\omega_0,\;\omega_1,\;\omega_3,\;\omega_4,\;\omega_5>0$, but $\omega_2$ is indefinite because $F_{k} - F_{V}$ is indefinite.
\begin{flalign}
	\label{eq66}
	\setlength{\belowdisplayskip}{6pt}
	\omega_{1}\omega_{5} & > \frac{2\alpha}{\eta E_{k}^{max}}\frac{\frac{\xi_{0}B_{0}}{\ln( 2 )}( {p_{k} + \vartheta_{k,c}} )}{d_{kv}^{2}R_{k}^{k\rightarrow v}( 1 + \frac{p_{k}\xi_{0}}{d_{kv}^{2}} )} & \notag \\
	& = \frac{2\alpha}{\eta E_{k}^{max}}\frac{\frac{\xi_{0}}{d_{kv}^{2}}( p_{k} + \vartheta_{k,c} )}{ln( 2 )( 1 + \frac{p_{k}\xi_{0}}{d_{kv}^{2}} )\log( 1 + \frac{p_{k}\xi_{0}}{d_{kv}^{2}} )} &\notag \\
	& > \frac{2\alpha}{\eta E_{k}^{max}}\frac{x + \frac{1}{8}x}{\ln( 2 )( 1 + x )\log( 1 + x )} > \frac{2\alpha}{\eta E_{k}^{max}}\,, &\tag{66}
\end{flalign}
where $0<x\ll 1$. However, the formula (65) still is indefinite. $P1.2.1$ is nonconvex optimization problem \cite{dai2018joint,boyd2004convex}.
%\begin{multline}
%  		
%\end{multline}

% you can choose not to have a title for an appendix
% if you want by leaving the argument blank
\section{PROOF OF THE LEMMA 2}
In $P1.2.1$, we use the extreme value discriminant in the multivariate function to check its convexity \cite{boyd2004convex}. Next, we calculate the sec-ond-order derivative of each variable in turn.

The second-order derivative of $U$ with respect to $\mathbf{F}$ is:
\begin{equation}
	\label{eq67}
	\frac{\partial^{2}U}{\partial\mathbf{F}^{2}} = - \frac{y_{km}\beta\varrho_{k}{\tilde{c}}_{k}}{{\ln\left( 2 \right)}T_{k}^{max}f_{km}^{3}\omega_{0}^{2}}( {\omega_{0} + \frac{y_{km}\beta{\tilde{c}}_{k}}{T_{k}^{max}f_{km}}} )\,. \tag{67}
\end{equation}

The second-order derivative of $U$ with respect to $\mathbf{F}$ and $\mathbf{P}$ is given by:
\begin{flalign}
	\label{eq68}
	\frac{\partial^{2}U}{\partial\mathbf{F}\partial\mathbf{P}} = \frac{\partial^{2}U}{\partial\mathbf{P}\partial\mathbf{F}} = & \frac{y_{km}\beta\varrho_{k}^{2}{\tilde{c}}_{k}}{{\ln\left( 2 \right)}T_{k}^{max}f_{km}^{2}\omega_{0}^{2}} &\notag \\
	& \times ( \frac{\alpha I_{k}}{\eta E_{k}^{max}R_{k}^{k\rightarrow v}} - \omega_{1}\omega_{3} )\,. &\tag{68}
\end{flalign}

The second-order derivative of $U$ with respect to $\mathbf{F}$ and $\boldsymbol{\varrho}$ is denoted as:
\begin{flalign}
	\label{eq69}
	\frac{\partial^{2}U}{\partial\mathbf{F}\partial\boldsymbol{\varrho}} =& \frac{\partial^{2}U}{\partial\boldsymbol{\varrho}\partial\mathbf{F}} =  \frac{y_{km}\beta{\tilde{c}}_{k}}{{\ln\left( 2 \right)}T_{k}^{max}f_{km}^{2}\omega_{0}^{2}} &\notag \\
	& \times ( \omega_{0} + \varrho_{k}( \frac{I_{k}}{R_{k}^{k\rightarrow v}}\omega_{1} + \frac{I_{k}}{R_{k}^{v\rightarrow m}} + \omega_{2} ))\,. &\tag{69}
\end{flalign}

Combining with (62), (63) and (64) in Lemma 1, the Hessian matrix of $P1.2.2$ is expressed as:
\begin{equation}
	\label{eq70}
	H(U) = \left[\begin{IEEEeqnarraybox*}[][c]{,c/c/c,}
		\vspace{3pt}
  		\frac{\partial^{2}U}{\partial\boldsymbol{\varrho}^{2}} & \frac{\partial^{2}U}{\partial\boldsymbol{\varrho}\partial\mathbf{P}} & \frac{\partial^{2}U}{\partial\boldsymbol{\varrho}\partial\mathbf{F}} \\ \vspace{3pt}
		\frac{\partial^{2}U}{\partial\mathbf{P}\partial\boldsymbol{\varrho}} & 	\frac{\partial^{2}U}{\partial\mathbf{P}^{2}} & \frac{\partial^{2}U}{\partial\mathbf{P}\partial\mathbf{F}} \\ \vspace{1pt}
		\frac{\partial^{2}U}{\partial\mathbf{F}\partial\boldsymbol{\varrho}} & 	\frac{\partial^{2}U}{\partial\mathbf{F}\partial\mathbf{P}} & \frac{\partial^{2}U}{\partial\mathbf{F}^{2}} \\  
	\end{IEEEeqnarraybox*}\right]\,. \tag{70}
\end{equation}
\setlength{\lineskip}{2.5pt}
\setlength{\lineskiplimit}{2.5pt}
Note that, $\frac{\partial^{2}U}{\partial\boldsymbol{\varrho}^{2}} = - \frac{1}{\ln\left( 2 \right)}( \frac{\frac{I_{k}\omega_{1}}{R_{k}^{k\rightarrow v}} + \omega_{2}}{\omega_{0}})^{2} < 0 $ is slightly changed with respect to (62) because $t_k^{k\rightarrow v}$ is different from $t_k^{k\rightarrow m}$ (same for $\omega_2$ and $\frac{\partial^{2}U}{\partial\boldsymbol{\varrho}\partial\mathbf{P}}$. However,  the conclusion is the consistent. According to the property of Hessian matrix \cite{sun2019approach}. The necessary and sufficient condition for $P1.2.2$ is that the matrix $H(U)$ is at least positive semi-definite. However, it can be obtained that $\frac{\partial^{2}U}{\partial\boldsymbol{\varrho}^{2}}\frac{\partial^{2}U}{\partial\mathbf{P}^{2}} - \frac{\partial^{2}U}{\partial\boldsymbol{\varrho}\partial\mathbf{P}}\frac{\partial^{2}U}{\partial\mathbf{P}\partial\boldsymbol{\varrho}}$ is indefinite based on Lemma 2. Thus, $P1.2.2$ is also non-convex.
% use section* for acknowledgment
\section*{Acknowledgment}
This work was supported in part by the National Natural Science Foundation of China (No. 62072475, No. 61772554).

% Can use something like this to put references on a page
% by themselves when using endfloat and the captionsoff option.
%\ifCLASSOPTIONcaptionsoff
%  \newpage
%\fi

% trigger a \newpage just before the given reference
% number - used to balance the columns on the last page
% adjust value as needed - may need to be readjusted if
% the document is modified later
%\IEEEtriggeratref{8}
% The "triggered" command can be changed if desired:
%\IEEEtriggercmd{\enlargethispage{-5in}}

% references section

% can use a bibliography generated by BibTeX as a .bbl file
% BibTeX documentation can be easily obtained at:
% http://mirror.ctan.org/biblio/bibtex/contrib/doc/
% The IEEEtran BibTeX style support page is at:
% http://www.michaelshell.org/tex/ieeetran/bibtex/
%\bibliographystyle{IEEEtran}
% argument is your BibTeX string definitions and bibliography database(s)
%\bibliography{IEEEabrv,../bib/paper}
%
% <OR> manually copy in the resultant .bbl file
% set second argument of \begin to the number of references
% (used to reserve space for the reference number labels box)
%\begin{thebibliography}{1}
%\includegraphics[width =0.96\linewidth]{images/Exp4_3.png}
\bibliographystyle{IEEEtran}
\bibliography{mybibfile}
\vspace{-10 mm}

\raggedbottom
%\end{thebibliography}

% biography section
% 
% If you have an EPS/PDF photo (graphicx package needed) extra braces are
% needed around the contents of the optional argument to biography to prevent
% the LaTeX parser from getting confused when it sees the complicated
% \includegraphics command within an optional argument. (You could create
% your own custom macro containing the \includegraphics command to make things
% simpler here.)
%\begin{IEEEbiography}[{\includegraphics[width=1in,height=1.25in,clip,keepaspectratio]{mshell}}]{Michael Shell}
% or if you just want to reserve a space for a photo:

% if you will not have a photo at all:
%\begin{IEEEbiographynophoto}{John Doe}
%Biography text here.
%\end{IEEEbiographynophoto}
%
%% insert where needed to balance the two columns on the last page with
%% biographies
%%\newpage
%
%\begin{IEEEbiographynophoto}{Jane Doe}
%Biography text here.
%\end{IEEEbiographynophoto}

% You can push biographies down or up by placing
% a \vfill before or after them. The appropriate
% use of \vfill depends on what kind of text is
% on the last page and whether or not the columns
% are being equalized.

%\vfill

% Can be used to pull up biographies so that the bottom of the last one
% is flush with the other column.
%\enlargethispage{-5in}

% that's all folks
\end{document}